\documentclass{sda-arxiv}
\usepackage[style=apa,natbib=true]{biblatex}
\usepackage{amssymb}
\usepackage{amsmath}
\usepackage{array}
\usepackage{tabularx}
\usepackage{booktabs}
\usepackage{algorithm}
\usepackage{algorithmic}
\usepackage{subcaption}
\let\cite\citep

\title{How Information Evolves: Stability-Driven Assembly and the Emergence of a Natural Genetic Algorithm}

\auth{
    Dan Adler \\\texttt{dan@danadler.com}
}

\abs{
Information can evolve as a physical consequence of non-equilibrium dynamics, even in the absence of genes, replication, or predefined fitness functions. We present Stability-Driven Assembly (SDA), a framework in which stochastic assembly combined with differential persistence biases populations toward longer-lived motifs. Assemblies that persist longer become more frequent and are therefore more likely to participate in subsequent interactions, generating feedback that reshapes the population distribution and implements fitness-proportional sampling, realizing evolution as a natural, emergent genetic algorithm (SDA/GA) driven solely by stability. We apply SDA/GA to chemical symbol space using SMILES fragments with recombination, mutation, and a heuristic stability function. Simulations show hallmark features of evolutionary search, including scaffold-level dominance, sustained novelty, and entropy reduction, yielding open-ended dynamics absent from equilibrium models with fixed transition rates. These results motivate an evolutionary ladder hypothesis where persistence-driven selection precedes genetic replication.
}

\begin{document}

\coverpage           %
\doublespacing       %


\section{Introduction}

One of the deepest questions in the origins of life research is how selection could operate before the advent of replication. Multiple theoretical frameworks have addressed how complexity might emerge from abiotic systems. The autocatalytic set theory shows how reaction networks can become self-sustaining \cite{kauffman1986autocatalytic, hordijk2011required}. Eigen's hypercycles \cite{eigen1979hypercycle} extend this idea to mutually reinforcing replicators, but still presuppose replication machinery. Fontana's algorithmic chemistry \cite{fontana1991algorithmic}, part of the broader field of Artificial Chemistries \cite{dittrich2001artificial,banzhaf2015artificial}, and later digital artificial life platforms such as \textit{Tierra} and \textit{Avida} \cite{ray1992tierra,adami1994avida} demonstrated open-ended dynamics in abstract symbolic systems. Banzhaf and Yamamoto \cite{banzhaf2015artificial} observe that genetic algorithms can be cast as Artificial Chemistries, with crossover and mutation acting as reactions on a population of strings, a framing that we will use in Section~\ref{sec:chemistry} of this paper.

More recently, Cronin's chemputation approach has automated chemical synthesis \cite{cronin2024chemputation}, showing how chemistry can be controlled as code, although only under the guidance of an external programmer. The GARD model of lipid assemblies \cite{segre2000compositional,markovitch2012universal} and protocell ABMs \cite{damer2015coupled} demonstrate how compositional inheritance can arise in specific chemistries. Each of these frameworks contributes valuable insight, yet they remain tied to particular substrates or assume replication.

Other perspectives emphasize general physical drivers of structure formation, including preferential attachment in networks \cite{barabasi1999emergence}, dissipative adaptation in nonequilibrium thermodynamics \cite{prigogine1977self, england2015dissipative}, and evolutionary dynamics in chemical reaction networks \cite{wu2009origin, nowak2006evolutionary}. Work on physical self-assembly has further shown that local interaction rules and constraints can generate modular structure and reusable subunits without explicit optimization or target states. Assembly theory formalizes this by quantifying complexity as the minimal depth of construction required to build a structure \cite{sharma2023assembly}, while constructor theory reframes physics in terms of which transformations are possible or impossible \cite{deutsch2013constructor}. Together, these frameworks provide powerful ways to characterize structure and complexity, but they largely remain descriptive: they do not explain how information-generating dynamics arise in real time through feedback between persistence, population statistics, and generative processes.

The phrase 'chemical evolution' is widely used in the literature to describe the progressive transformation of molecules into more complex structures under prebiotic conditions. Yet the term is often applied descriptively rather than mechanistically, leaving open the question of what underlying principle imparts directionality or selection to such processes. In particular, while many models demonstrate how complexity can arise in specific chemistries, a general account of how chemical systems move from undirected reaction networks to cumulative evolutionary dynamics has remained elusive.

Biological evolution depends on variation and inheritance, but both presuppose molecular mechanisms capable of copying information. However, long before such machinery existed, physical and chemical systems already exhibited differences in persistence: some structures survived longer than others under given conditions. The central question is whether these persistence imbalances, acting alone, are sufficient to generate selection-like dynamics and drive the emergence of complexity.

In inanimate systems, patterns persist according to their stability under prevailing conditions. Diamonds outlast graphite under high pressure, and certain molecular motifs are more resilient in given chemical environments \cite{ruizmirazo2014}. Here, we use motif to mean a distinct pattern or, in chemical contexts, a distinct species. Such differences in longevity create implicit selection: stable structures naturally accumulate, biasing the population distribution. This principle of stability-driven selection may provide a bridge between nonliving matter and evolving biospheres \cite{kauffman1993origins, hordijk2012autocatalytic, nghe2015prebiotic}.

Crucially, the persistence imbalances relevant to prebiotic and abiotic evolution are not small perturbations, but span many orders of magnitude. Across physics and chemistry, binding energies, decay rates, and structural reinforcement mechanisms generate lifetime differences ranging from fleeting quantum states to structures that persist for geological or cosmological timescales. These disparities are not incidental: they arise generically from symmetry, energetic minima, spatial organization, and hierarchical assembly. As a result, even weak biases in persistence, when integrated over time and coupled to continual stochastic assembly, can dominate population statistics. In such settings, the probability distribution over structures is not sampled uniformly but becomes progressively skewed toward long-lived motifs, producing effective directionality and entropy reduction without invoking replication, templating, or externally imposed fitness criteria.

In this work, we show that persistence imbalances alone are sufficient to implement an evolutionary search. Specifically, we demonstrate that when stochastic assembly is coupled to differential persistence in an open, replenished system, the resulting feedback on the population distribution necessarily realizes fitness-proportional sampling. This establishes evolution as a natural genetic algorithm: an optimization process that is neither externally programmed nor combinatorial, but arises generically from persistence-weighted dynamics in nonequilibrium systems.

\section{Stability-Driven Assembly (SDA)}

Stability-Driven Assembly (SDA) was previously introduced \cite{adler_sda} as a general non-equilibrium framework in which persistence imbalances bias the accumulation of structure over time . The present work does not revisit the full theoretical development of SDA, but instead draws on its minimal formal structure to analyze the resulting population dynamics. Accordingly, this section summarizes only those elements of SDA necessary to establish how persistence-weighted sampling reshapes probability distributions and induces evolutionary search.

A Stability-Driven Assembly (SDA) system consists of base elements $A, B, C, \dots$ that recursively combine into compounds represented as
strings of unbounded length. The persistence of each compound is determined by its
stability $S$: a compound with $S=30$ remains for 30 generations before elimination, while the patterns with $S=1$ vanish after a single generation. 
Expired motifs are removed without explicit reverse reactions and base
elements are replenished at a constant rate each generation, ensuring
continued exploration of assembly space. This setup parallels continuous-flow 
stirred tank reactors (CFSTR) \cite{fogler1999chemical}, where reactants are
continuously supplied to maintain non-equilibrium conditions. More formally, An SDA system is defined as a tuple $(E, P, S, R, I)$ where:
\begin{itemize}
   \item[] $E = \{e_1, e_2, \ldots, e_n\}$ is a finite set of base elements
   \item[] $P$ is the set of all possible patterns formed by concatenating elements from $E$ and existing patterns
   \item[] $S: P \rightarrow \mathbb{Z}^{+}$ is a stability function mapping each pattern to a positive integer representing its lifetime
   \item[] $R: E \rightarrow \mathbb{Z}^{+}$ is a replenishment function that specifies how many copies of each base element are added per generation
   \item[]  $I \in \mathbb{Z}^{+}$ is the number of interactions allowed per generation
\end{itemize}
The pattern interaction operation, denoted by $\oplus$, is defined as a string concatenation. When patterns $p_1$ and $p_2$ interact, they form a pattern $p_1 \oplus p_2$.

\begin{algorithm}[H]
\caption{Stability-Driven Assembly (SDA)}
\begin{algorithmic}[1]
\REQUIRE Base elements $E$, stability $S$, replenish $R$, interactions $I$, generations $T$
\STATE Initialize population with $E$
\FOR{$t=1$ to $T$}
  \STATE Remove expired patterns; add $R(e)$ for all $e\in E$
  \FOR{$i=1$ to $I$}
    \STATE Sample $p_1,p_2$ from population
    \STATE $p\leftarrow p_1\oplus p_2$; set expiration $t+S(p)$; add $p$ to population
  \ENDFOR
\ENDFOR
\end{algorithmic}
\end{algorithm}

The probability of selecting a pattern $p$ for interaction is proportional to its frequency in the population. This creates a feedback mechanism: patterns with higher stability persist longer, becoming more abundant, which increases their probability of participating in interactions.

The original SDA formulation emphasizes string concatenation as the sole
pattern-forming operation. In what follows, we generalize the interaction 
to a recombination operator with an optional mutation step, and then examine 
whether the core properties of SDA are preserved under this extension. 
We refer to this generalized version as \textit{SDA/GA}, as the dynamics will be shown to 
be equivalent to those of a genetic algorithm, while retaining 
persistence as the implicit fitness measure. Instead of the SDA concatenation:
\[
p_{\text{new}} = p_1 \oplus p_2
\]
we construct new patterns by recombining substrings of both patterns:
\[
p_{\text{new}} = \mathrm{Recombine}(p_1, p_2)
\]
optionally followed by \emph{single–site mutation} (localized stochastic modification of an existing pattern, such as insertion, deletion, or substitution at a single site) applied with probability $\mu$:
\[
p_{\text{new}} =
\begin{cases}
\mathrm{Mutate}(p_{\text{new}}) & \text{with probability } \mu,\\
p_{\text{new}} & \text{with probability } 1-\mu
\end{cases}
\]

These changes do not alter the selection mechanism: persistence $S(p)$ still 
determines the residence time and thus induces the same frequency-weighted sampling characteristic of roulette–wheel selection.

\subsection{Population and Stability Distributions}

SDA dynamics are captured by two complementary distributions. 
The \textit{population distribution} tracks the relative abundance of each pattern:
\begin{equation}
P_t(p) = \frac{N_t(p)}{\sum_{q \in P} N_t(q)},
\end{equation}
where $N_t(p)$ is the count of the pattern $p$ in generation $t$. 
The \textit{stability function}, $S: P \rightarrow \mathbb{Z}^+$, assigns each pattern a characteristic lifetime. 
In chemical implementations, these lifetimes could in principle be derived from binding energies, activation barriers, or other domain-specific stability measures.

\subsection{Population-level Analysis of SDA Dynamics}
\label{sec:pattern-evolution}

In \cite{adler_sda} we derived a simple persistence–creation update rule where 
recombination was modeled as deterministic concatenation. This took the form
\begin{equation}
\label{eq:create-term-simple}
\mathrm{Create}_t(p) = \sum_{(q,r)\to p} P_t(q)\,P_t(r),
\end{equation}
where $(q,r)\to p$ denotes parent pairs that yield $p$ by concatenation. 
Together with the persistence term
\begin{equation}
\label{eq:persist-term-simple}
\mathrm{Persist}_t(p) = P_t(p)\left(1-\frac{1}{\overline{R}_t(p)}\right),
\end{equation}
where $\overline{R}_t(p)$ denotes the mean remaining lifetime of all active
instances of $p$ at time $t$. This gives a baseline recursive update. To handle recombination more generally, we now allow a parent pair $(q,r)$ 
to produce not just one deterministic child but a distribution of possible
offspring. We capture this by introducing a recombination-mutation kernel
$K_t(p\mid q,r)$, which gives the probability that $q$ and $r$ produce
offspring $p$ at time $t$:
\begin{equation}
\label{eq:create-term-kernel}
\mathrm{Create}_t(p) = \sum_{q,r\in P} P_t(q)\,P_t(r)\,K_t(p\mid q,r),
\qquad \sum_{p\in P} K_t(p\mid q,r)=1~\forall~q,r.
\end{equation}
The persistence term remains as in Eq.~\ref{eq:persist-term-simple}. Intuitively, $K_t$ is just a lookup table of possible children for each 
parent pair. In the deterministic concatenation case, it reduces to 
$K_t(p\mid q,r)=\mathbf{1}\{p=q\oplus r\}$, so each pair produces exactly 
one child. In recombination, $K_t$ may spread the probability mass across 
several outcomes: most weight might go to $q\oplus r$, but the smaller weight 
can go to shorter or mutated variants. This makes explicit how mutation 
and recombination introduce variation, while preserving the same update 
structure as before. The updated population distribution is then:
\begin{equation}
\label{eq:full-ba-update}
P_{t+1}(p) = \frac{\mathrm{Persist}_t(p) + \mathrm{Create}_t(p)}
{\sum_{p' \in P} \big[\mathrm{Persist}_t(p') + \mathrm{Create}_t(p')\big]}.
\end{equation}

Thus, the kernel-based formulation extends the earlier model in a minimal
way: concatenation is a special case, and recombination with mutation is
just a richer offspring distribution. This does not alter the persistence-driven drift, which remains governed entirely by stability. The kernel instead modifies the structure of the creation term, enriching the range of variants without changing the underlying bias toward stability.

\subsection{Entropy Dynamics}

The Shannon entropy of the population at time $t$ is defined as
\begin{equation}
H(P_t) = - \sum_{p \in P} P_t(p) \log_2 P_t(p).
\end{equation}
Changes in entropy reflect the balance between novelty introduced by creation and order enforced by persistence.
\begin{equation}
\Delta H = H(P_{t+1}) - H(P_t).
\end{equation}
To quantify this balance, we define
\begin{equation}
\alpha = \frac{\sum_p \mathrm{Persist}_t(p)}
{\sum_p \mathrm{Persist}_t(p) + \sum_p \mathrm{Create}_t(p)}.
\end{equation}
When $\alpha \to 1$, stability dominates and entropy tends to decrease; when $\alpha \to 0$, creation dominates and entropy tends to increase.  
A first-order approximation expresses the entropy change as a weighted combination of these contributions:
\begin{equation}
\Delta H \approx (1 - \alpha)\,\Delta H_{\text{create}} + \alpha\,\Delta H_{\text{persist}},
\end{equation}
where $\Delta H_{\text{create}}$ is typically positive and $\Delta H_{\text{persist}}$ negative.  The interaction of these opposing forces explains the oscillatory entropy trajectories sometimes observed in simulation, as phases of innovation alternate with phases of stabilization.

\subsection{Connection to Continuous and Thermodynamic Formulations}

The discrete dynamics described above admit a master equation formulation \cite{adler_sda} in which the time evolution of $P_t(p)$ is written as a balance of creation and persistence terms over the pattern space. This maps to a nonlinear Fokker--Planck (McKean--Vlasov) system in which the drift functional depends self-consistently on the evolving distribution \cite{mckean1966, villani2009}. An equivalent thermodynamic formulation casts pattern creation as the energy-driven crossing of activation barriers that encode information into persistent structures, and pattern decay as Landauer-style erasure of that information \cite{landauer1961irreversibility, bennett1973logical}. A formal development of these connections, including the construction of a metric pattern-space and the derivation of the thermodynamic form, is left for forthcoming work.

\subsection{Symbolic Simulation Model Parameters}

All symbolic SDA and Unconstrained simulations below used a common set of parameters. The base elements are $A$, $B$, $C$.
Each generation included
a replenishment of 5 base elements and 100 random pairwise
interactions. 

In the unconstrained simulations, all the resulting patterns had the same stability of 1 generation, so the dynamics reduced to random search: no patterns persist across generations, no memory accumulates, and no bias arises to skew the population distribution which remains uniform. 

For the SDA simulations, 
stability values were assigned to patterns according to the
function:

\begin{equation}
S(p) =
\begin{cases}
100, & \text{if } p = \text{ABCABA} \\
50, & \text{if } p \in \{\text{ABA, ABC}\} \\
30, & \text{if } p \in \{\text{AB, BC}\} \\
1, & \text{otherwise}.
\end{cases}
\end{equation}

These parameters were fixed in the symbolic experiments below, ensuring that the observed differences arose from the interventions rather than changes in
the baseline setup.

\begin{figure}[H]
    \centering
    \includegraphics[width=0.6\textwidth]{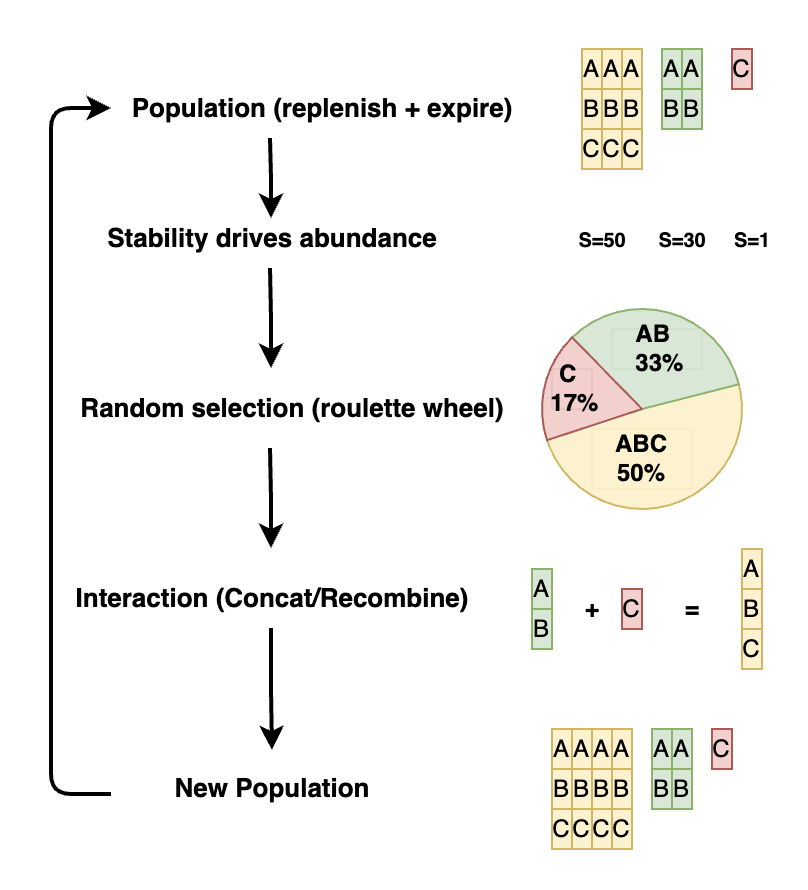}
    \caption{Symbolic SDA/GA loop in ABC space. 
Base elements are replenished while unstable motifs expire, yielding a population skewed by stability ($S=50,30,1$). 
Roulette-wheel sampling proportional to abundance selects motifs for interaction (concatenation in SDA, recombination in GA), 
generating new assemblies. 
This feedback of \textit{stability} $\rightarrow$ \textit{persistence} $\rightarrow$ \textit{population skew} produces emergent fitness-proportional selection without an explicit fitness function.}
    \label{fig:sda-loop}
\end{figure}

Figure~\ref{fig:sda-loop} illustrates how these parameters operate within the
symbolic SDA/GA loop. Relenishment and expiration define the active population, 
stability values skew motif abundance, and roulette-wheel sampling drives
biased parent selection. Interactions (concatenation or recombination) then
produce new motifs that re-enter the pool, closing the feedback cycle. 
This diagram highlights how persistence alone induces effective selection
pressure without an explicit fitness function.

\subsection{Simulation Results}

\begin{figure}[H]
    \centering
    \includegraphics[width=1\textwidth]{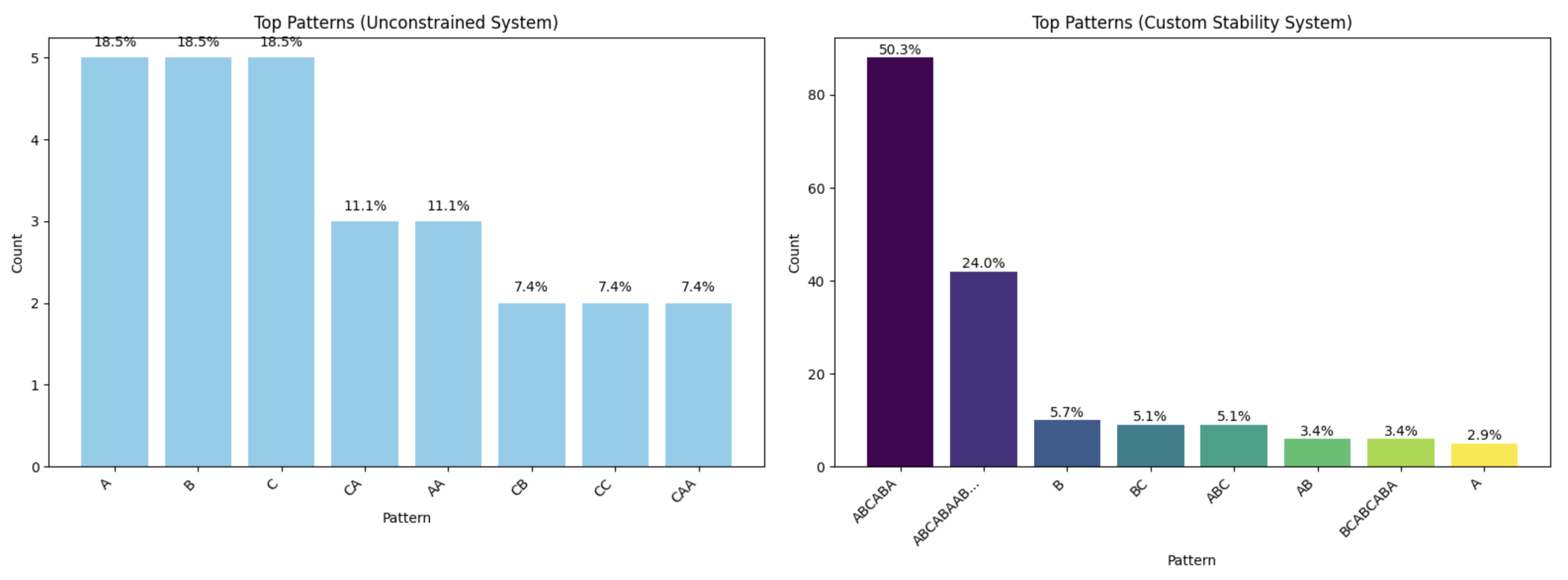}
\caption{Pattern distribution in the SDA system with concatenation.
\textit{Left:} with no persistence bias yields a
near-uniform distribution.
\textit{Right:} stability-driven persistence produces dominance of a small set
of high-stability motifs.}
    \label{fig:concat-patterns}
\end{figure}

\begin{figure}[H]
    \centering
    \includegraphics[width=1\textwidth]{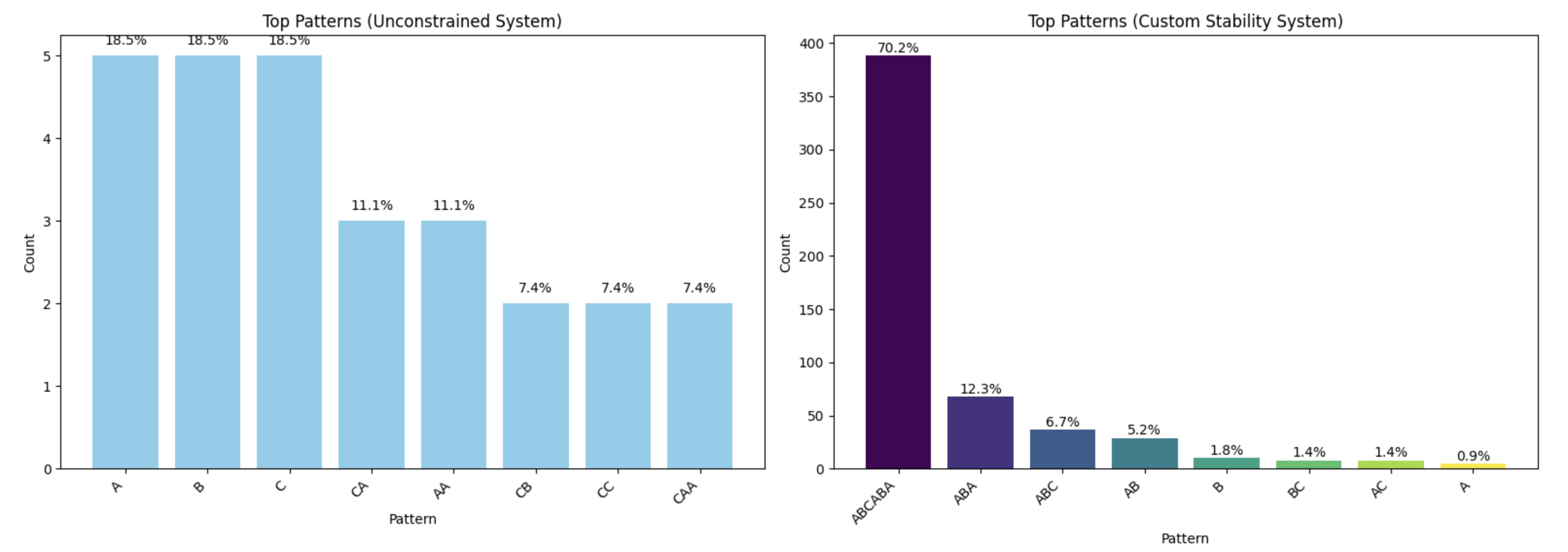}
\caption{Pattern distribution in the generalized SDA/GA system with recombination.
\textit{Left:} with no persistence bias yields a
near-uniform distribution.
\textit{Right:} persistence-driven selection yields dominant stable motifs with
a broader low-frequency tail maintained by recombination.}
    \label{fig:ga-patterns}
\end{figure}

We begin by comparing the final pattern distributions produced by the original SDA  with concatenation \cite{adler_sda} and the generalized SDA/GA with recombination. Both systems show a strong deviation from the unconstrained case: instead of a uniform spread of short strings, high-stability motifs dominate the population. In the SDA system, concatenation drives the emergence of long-repeated motifs such as \texttt{ABCABA}, which accumulate a lot of the probability mass. In the GA variant based on recombination, the same dominant motifs appear, but the distribution shows a heavier tail: many low-frequency variants persist along the stable core (Figures~\ref{fig:concat-patterns},~\ref{fig:ga-patterns}). This broader spectrum reflects the ability of recombination to continuously inject mosaic offspring and maintain a pool of rare types, while pure concatenation channels more strongly into layered repeats.

\begin{figure}[H]
\centering
\subfloat[\centering Concatenation-based SDA]{
    \includegraphics[width=0.45\textwidth]{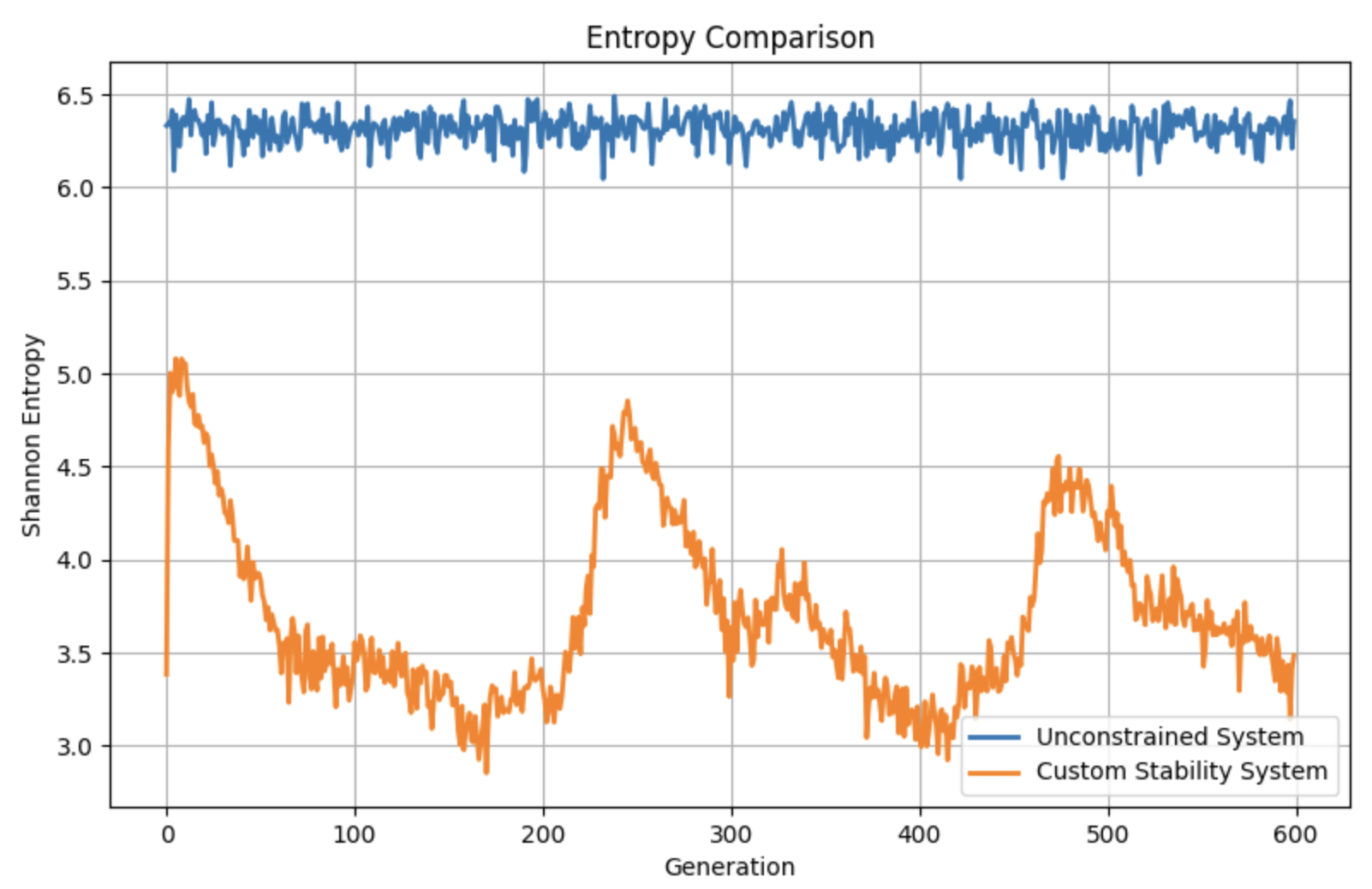}
    \label{fig:concat-entropy}
}
\hfill
\subfloat[\centering Recombination-based SDA/GA]{
    \includegraphics[width=0.45\textwidth]{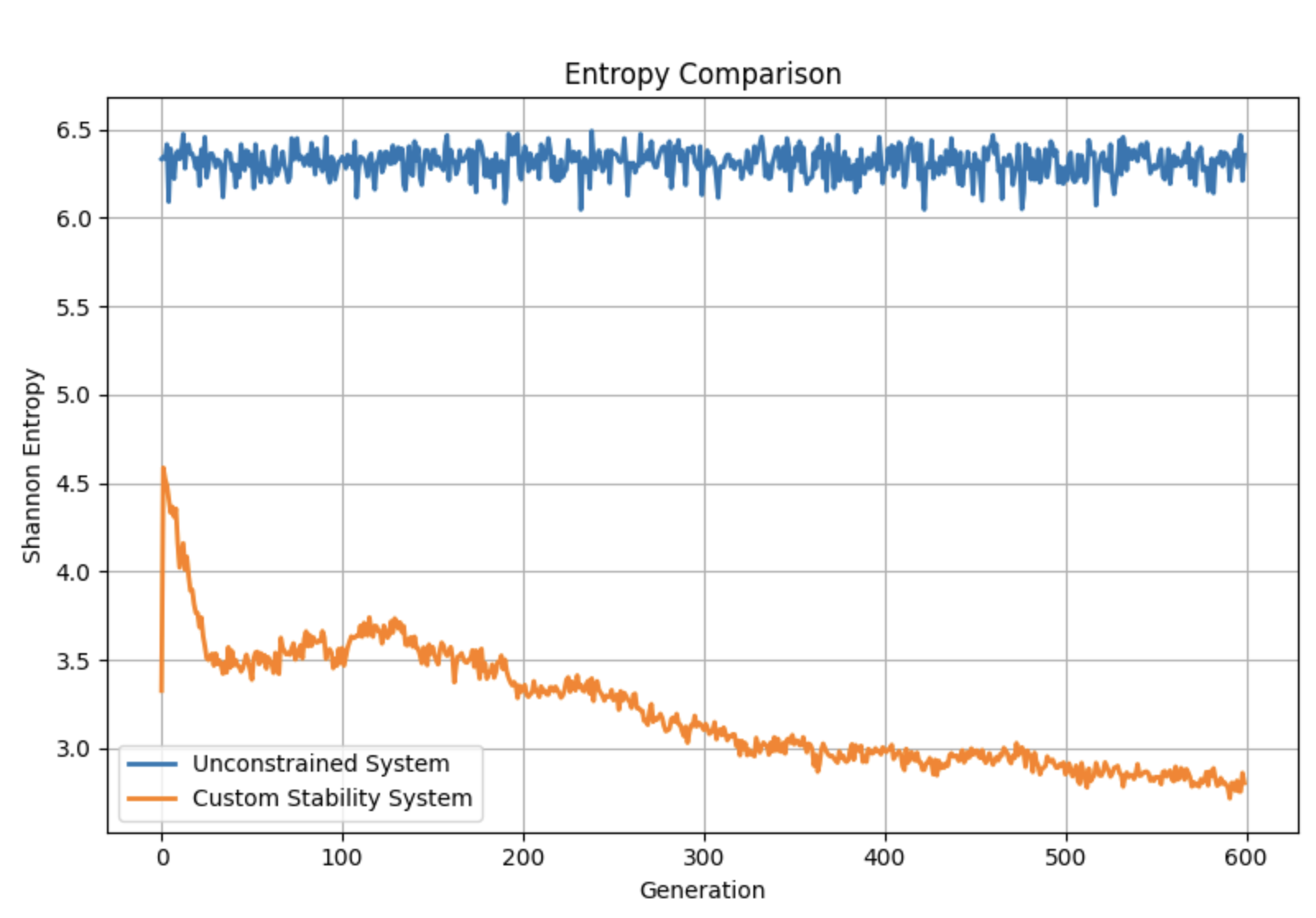}
    \label{fig:ga-entropy}
}
\caption{Entropy dynamics under different operators.
In both panels, the upper blue curve shows the unconstrained control without
persistence bias.
(a) Concatenation-based SDA exhibits oscillatory boom–bust entropy cycles due
to synchronized expiration of long motifs.
(b) Recombination-based SDA/GA shows smoother entropy decline, as recombination
desynchronizes expirations.}
\label{fig:entropy-comparison}
\end{figure}

In terms of entropy dynamics, both systems show a clear reduction in Shannon entropy ($H(P_t) = - \sum_{p \in P} P_t(p) \log_2 P_t(p)$) of
the pattern distribution in generation t measured in bits relative to the unconstrained baseline, confirming the emergence of order and the presence of selection pressure (Figures~\ref{fig:concat-entropy},~\ref{fig:ga-entropy}). Importantly, this occurs without any explicit fitness-proportional selection rule in the algorithm: roulette-wheel selection emerges intrinsically from persistence, since patterns with longer lifetimes are overrepresented and thus more likely to be sampled for further interactions.

In concatenation-based SDA, while the average entropy decreases from $\sim6$ bits to $\sim4$ bits, its trajectories often exhibit oscillations. These arise because concatenation tends to build large, synchronized cohorts of similar long motifs that expire at nearly the same time. When such a cohort collapses, replenished base elements briefly increase diversity and entropy before new dominant motifs emerge, creating a characteristic boom–bust cycle. In contrast, in the SDA/GA based on recombination, the entropy decreases more smoothly to $\sim3$ bits without oscillations. Recombination produces mosaic offspring with staggered lifetimes, desynchronizing expirations, and damping collective turnover. The result is a more monotonic entropy collapse toward a skewed distribution anchored by stable motifs.

Together, these results demonstrate that SDA robustly yields emergent selection pressure and entropy reduction under both concatenation and recombination operators. The choice of interaction operator primarily shapes the dynamical form of convergence: oscillatory cycles under concatenation versus smooth decline under recombination.

\begin{figure}[H]
\centering
\subfloat[\centering Concatenation-based SDA]{
    \includegraphics[width=0.45\textwidth]{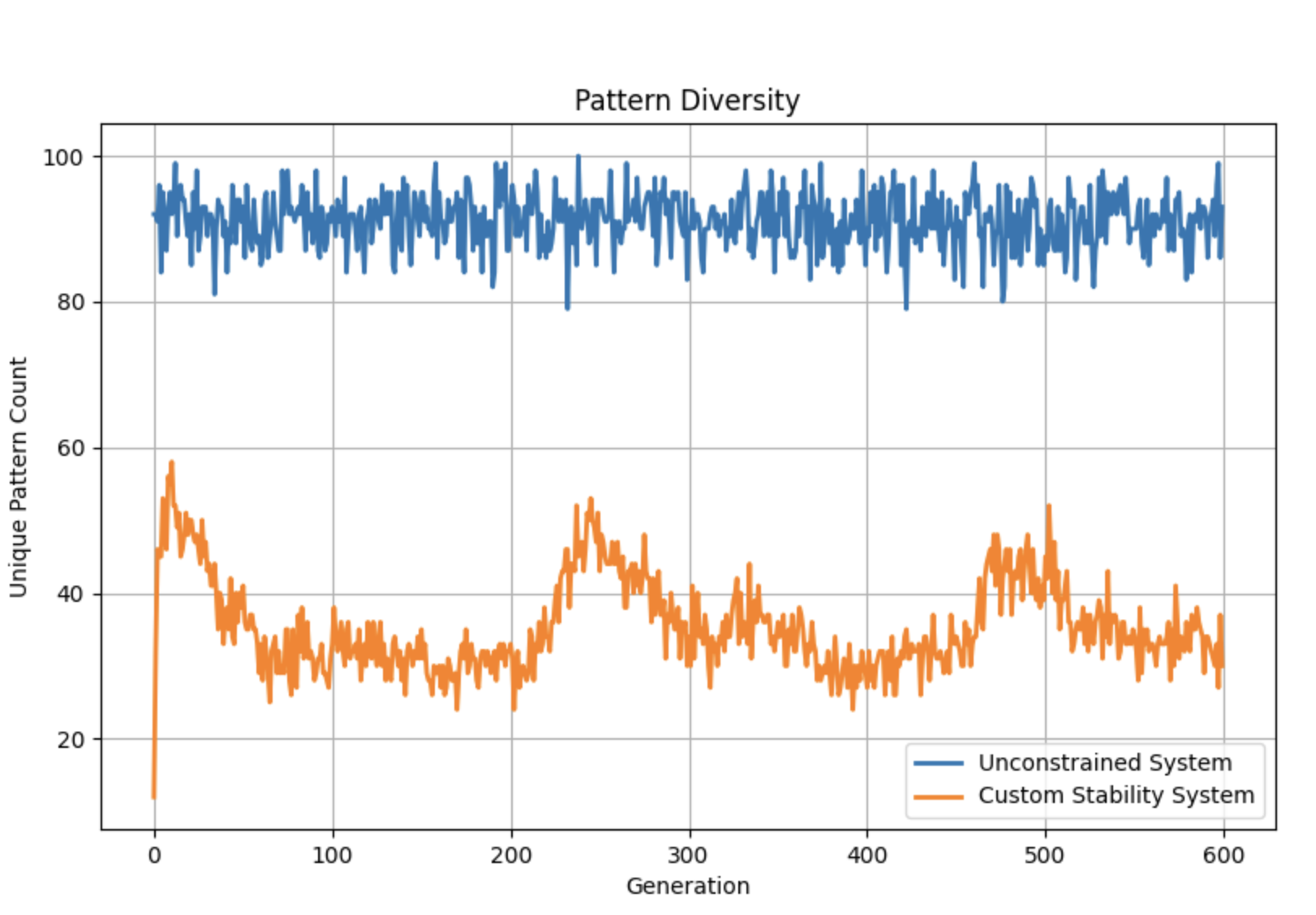}
    \label{fig:concat-diversity}
}
\hfill
\subfloat[\centering Recombination-based SDA/GA]{
    \includegraphics[width=0.45\textwidth]{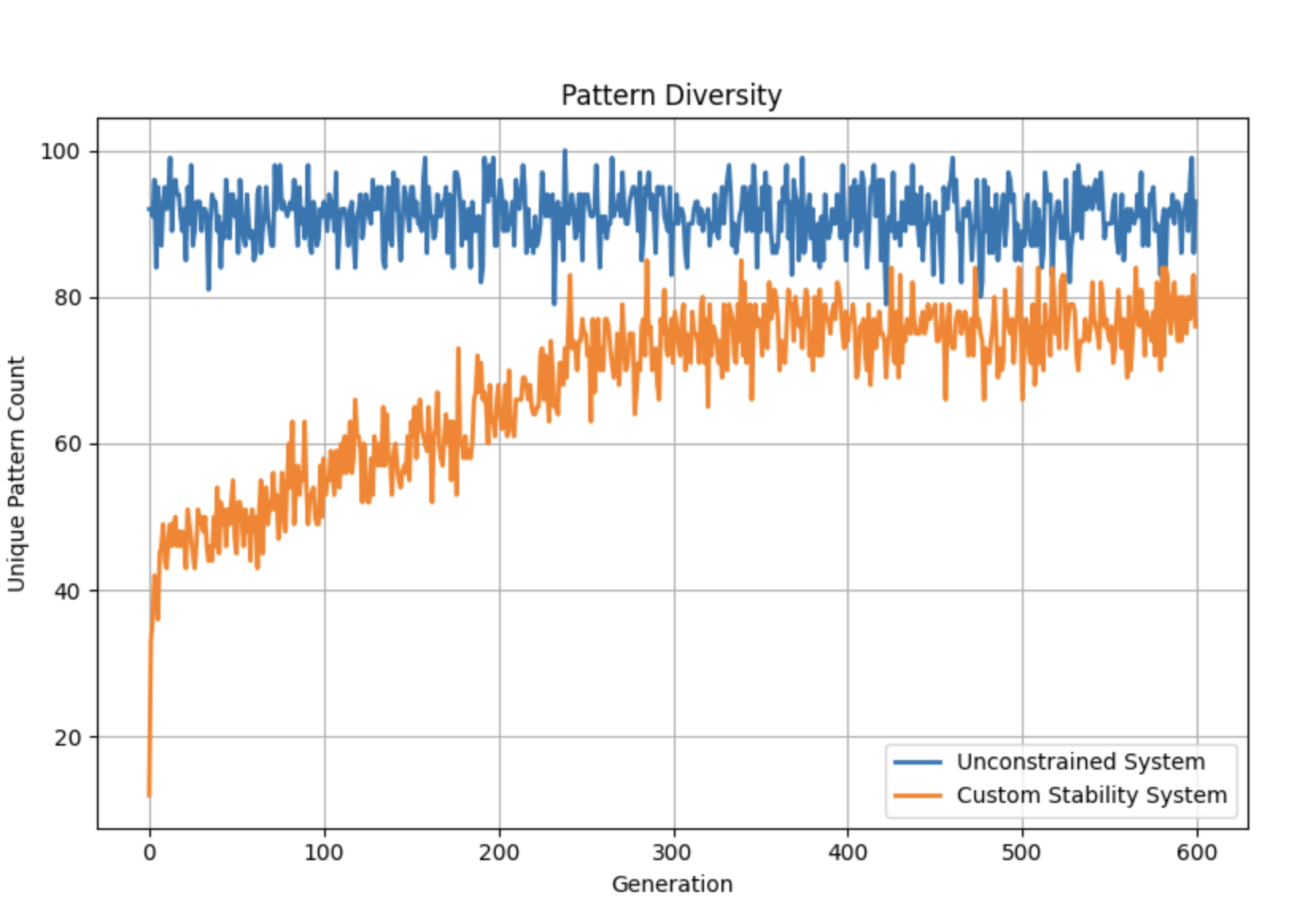}
    \label{fig:ga-diversity}
}
\caption{Pattern diversity under different operators.
In both panels, the upper blue curve shows the unconstrained control without
persistence bias.
(a) Concatenation-based SDA exhibits oscillations reflecting synchronized motif
turnover.
(b) Recombination-based SDA/GA shows steadily increasing diversity, indicating
continuous generation of low-frequency variants.}
\label{fig:diversity-comparison}
\end{figure}

The pattern diversity results in Figures~\ref{fig:concat-diversity} and~\ref{fig:ga-diversity} illustrate the exploration–exploitation tradeoff. Concatenation-based SDA exhibits intermittent exploration, with diversity rising during replenishment and collapsing as dominant motifs take over, producing boom–bust dynamics. In contrast, recombination-based SDA/GA maintains broad exploration, with diversity approaching unconstrained levels even as entropy declines. Most variants remain low-frequency, but their persistence reflects continuous novelty injection. Thus, SDA/GA naturally combines the exploitation of stable motifs with ongoing exploration of alternatives, a hallmark of genetic algorithms, without an explicit fitness function.

We also tested the effect of introducing low-probability single-site mutations during recombination. 
These mutations serve as a simple model of stochastic perturbations such as copying errors or
diffusion events. As expected, mutation did not qualitatively alter the dynamics: entropy
collapsed and high-stability motifs remained dominant. Mutations did not disrupt the overall reduction in entropy, confirming that the emergent selection mechanism is robust to stochastic noise.

\section{Genetic Algorithms and Natural Genetic Algorithms}

The study of genetic algorithms (GAs) has a long history in computer science and optimization. 
Holland’s foundational work \cite{holland1975adaptation} introduced the idea of using crossover, 
mutation, and selection to evolve solutions to computational problems. Later, Goldberg 
\cite{goldberg1989genetic} popularized and formalized GAs as practical optimization tools, 
particularly for engineering and combinatorial search \cite{adler1993marriage}. Building on this tradition, Koza \cite{koza1992genetic} extended the paradigm to genetic programming (GP), in which whole 
program trees rather than fixed-length strings evolve under crossover and mutation. GP has 
been especially influential in symbolic regression and in evolving nontrivial structures such 
as circuits, strategies, and controllers.

A useful concept in this literature is that of a schema: a partially specified pattern or scaffold shared by many individuals, which effectively acts as a reusable template during recombination. Schemata constrain the search by biasing variation toward combinations that preserve common substructures, improving efficiency without prescribing a specific outcome. Importantly, such constraints do not force convergence toward a predefined target; they shape the space of accessible variants while leaving the evolutionary process open-ended. As shown below, stability-driven assembly naturally gives rise to chemically meaningful schemata, enabling efficient exploration through shared scaffolds without imposing goal-directed optimization.

In classical GA and GP, selection is driven by an explicit fitness function supplied by the programmer. Candidate solutions are evaluated against a predefined objective, assigned fitness values, and sampled accordingly, so the selective pressure is imposed externally by the solution designer. In contrast, in SDA/GA systems there is no externally specified fitness function or selection operator. Instead, selection emerges intrinsically from persistence: each pattern has a stability S(p) that determines its lifetime, and patterns that persist longer naturally become more frequent and more likely to participate in further recombination. This persistence-weighted feedback implements fitness-proportional sampling without computing or prescribing a fitness function. In simple string models, stability can be assigned by hand for clarity, but in realistic settings it is determined by the environment itself: by thermodynamics, kinetics, structural constraints, and operating conditions in chemistry, or by ecological context in biology. Thus, while classical GAs supply fitness explicitly, SDA/GA derives fitness from environmental interaction, framing evolutionary search as an emergent consequence of stability, probability, and feedback rather than an externally imposed algorithmic goal.

A well-known illustration of genetic algorithms is Dawkins’ “weasel” program \cite{dawkins1986blind}, in which random sequences evolve toward a fixed target Shakespearean phrase through repeated variation and selection. While often cited to show that cumulative selection outperforms random search, the example relies on an externally defined target and an explicit fitness function measuring distance from that target, effectively implementing supervised optimization. This target-driven design contrasts with SDA/GA, which does not optimize toward any predefined objective: selection arises endogenously from persistence and feedback in an open system, yielding open-ended evolutionary dynamics rather than convergence to a specified solution.

An analogy closer to SDA is jazz improvisation \cite{adler2025jazz}. 
Musicians explore an open-ended space of motifs within a predefined harmonic space
without a fixed target. Motifs that resonate are repeated, varied, and recombined, while others fade. Musical structure thus emerges from this stochastic but biased search. 
Similarly, SDA explores assembly space, persistence bias amplifies
stable motifs, and novelty accumulates without a prespecified goal. 
Both processes discover order by letting persistence guide variation, 
illustrating how information can emerge from feedback rather than
external targets.

\section{Application to Organic Chemistry}
\label{sec:chemistry}

The abstract SDA/GA framework can be instantiated in the chemical symbol space to model
the spontaneous evolution of molecular populations. In this setting, the base elements
$E$ are drawn from the atomic alphabet (C, O, N, H, etc.), and the interaction operator
$\oplus$ is instantiated as recombination and mutation of molecular fragments. 

Genetic algorithms and genetic programming have long been applied in chemistry and 
cheminformatics, particularly for de novo molecular design and drug discovery 
\cite{brown2004ga,jensen2019ga,yoshikawa2018ga}. These approaches typically 
represent molecules as graphs or strings and apply crossover and mutation operators 
guided by an externally defined fitness function, such as binding affinity, 
drug-likeness, or synthetic accessibility. Fink and Reymond’s construction of the
GDB-11 database \cite{fink2007gdb11} demonstrated the sheer size of chemically valid
search space, generating over 26 million molecules with up to 11 atoms and over
110 million stereoisomers. However, only a small fraction of these compounds occur in
public databases, underscoring both the vastness of chemical possibility and the impracticality of a uniform or ergodic search in nature. This observation motivates the use of
search strategies that inherently bias exploration toward persistent and chemically
stable motifs. Whereas traditional GA/GP methods rely on explicit, human-specified 
fitness functions to impose such bias, the SDA/GA framework does so intrinsically by
embedding stability into persistence, allowing selection pressure to emerge directly
from environmental constraints such as valence rules, steric feasibility, and
thermodynamics.

It is important to emphasize that the present work is not aimed at drug discovery
or molecular optimization applications, which have been the traditional focus of
genetic algorithms and genetic programming in chemistry. Instead, our goal is
conceptual: to hypothesize how nature itself may have acted as a 'natural genetic algorithm', using stability-driven persistence as the implicit fitness function to
non-ergodically explore the astronomical chemical space revealed by studies such
as GDB-11. Within this perspective, the emergence of a biosphere can be viewed as
the outcome of stability-biased sampling: persistent motifs accumulate, dominate, 
and recombine, gradually transforming an unconstrained chemical universe into a
structured, evolving system.

The mapping in Table~\ref{tab:chem-sda-operators} illustrates how common classes of organic reactions can be interpreted within the SDA/GA framework. Substitution, reduction and oxidation reactions all correspond to \textit{mutations}, since they alter functional groups while preserving the underlying scaffold. Addition reactions may play either role: the attachment of a small atom or group is best regarded as mutation, whereas the joining of two larger fragments constitutes recombination.  

Acid-base reactions are a particularly simple but important form of mutation. Protonation and deprotonation cycles alter charge states and stability without changing the covalent backbone, yet they have profound effects on persistence in different environments. Similarly, isomerization represents a mutational step in which the connectivity or geometry is reshuffled, sometimes uncovering stability differences that bias persistence even though no atoms are gained or lost.  

Polymerization, including peptide bond formation, provides a canonical example of recombination. Here, motifs are linked through edge-biased joining to form larger assemblies. Once formed, such chains introduce new levels of persistence and complexity, laying the groundwork for the higher rungs of the evolutionary ladder. Fragmentation and dissociation serve as the inverse of recombination, redistributing persistence across smaller motifs.  

This mapping shows that the SDA/GA operators of mutation and recombination are not abstract inventions but correspond directly to well-established categories of chemical reactivity. This correspondence grounds the abstraction in chemical reality and illustrates how stability-driven selection could operate in real chemical networks without requiring additional operators beyond those already available to organic chemistry.

\begin{table}[t]
\caption{Representative mappings between classical organic reactions and SDA operators.}
\label{tab:chem-sda-operators}
\centering
\begin{tabularx}{\textwidth}{l X X}
\toprule
\textbf{Reaction Type} & \textbf{Representative Example} & \textbf{SDA Analogy} \\
\midrule
Substitution / Redox 
& R--CHO $\leftrightarrow$ R--CH$_2$OH / R--COOH 
& Mutation: functional group or bond order change \\

Addition 
& R--C=O + R$'$X $\rightarrow$ R--C(OH)X 
& Recombination (full fragment) or mutation (small group) \\

Acid--Base 
& R--NH$_2$ + H$^+$ $\leftrightarrow$ R--NH$_3^+$ 
& Mutation: reversible protonation \\

Polymerization 
& Amino acids $\rightarrow$ peptides 
& Recombination at reactive edges \\

Isomerization 
& Keto--enol tautomerism 
& Mutation: structural rearrangement \\

Fragmentation 
& Ester hydrolysis (R--COOR$'$ $\rightarrow$ R--COOH + R$'$OH) 
& Inverse recombination; persistence splits \\

\bottomrule
\end{tabularx}
\end{table}

In classical organic chemistry, reaction types are usually taught as synthesis pathways, 
each requiring defined reagents, catalysts, and conditions. The SDA framework, on the other hand, abstracts these transformations into population-level operators. 
Rather than focusing on the mechanistic sequence of steps, SDA emphasizes how
persistence and recombination biases determine which motifs accumulate over time. 
In this way, SDA complements the traditional synthesis perspective by revealing
how macro-level selection effects emerge from the distribution of possible
transformations.  

\begin{figure}[H]
\centering
\subfloat[\centering Mutation]{
  \includegraphics[width=5cm]{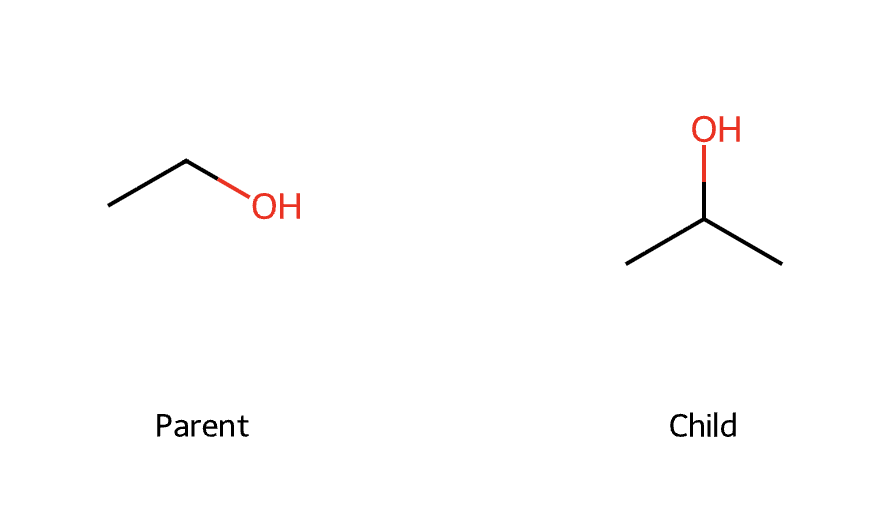}}
\hspace{0.4cm}
\subfloat[\centering Recombination]{
  \includegraphics[width=7cm]{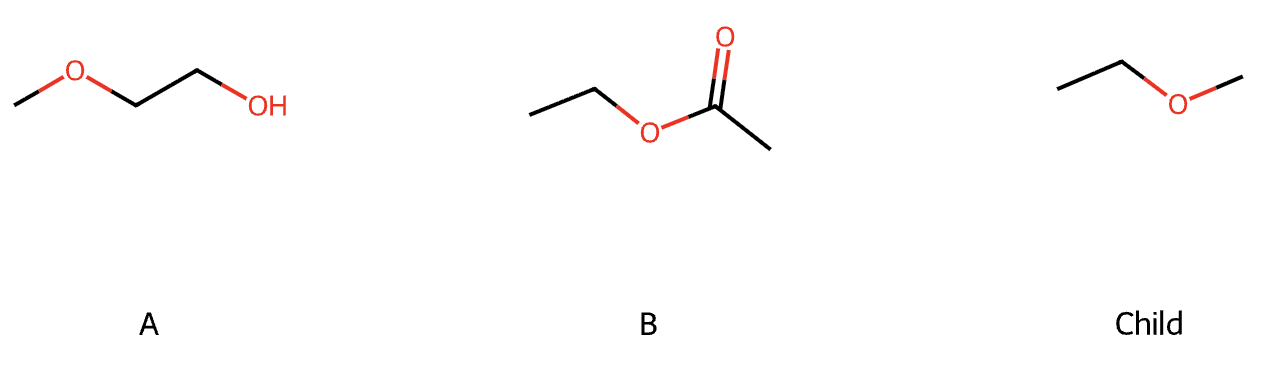}}
\caption{Examples of SDA operators in chemical form. (\textbf{a}) Mutation: ethanol CCO $\to$ isopropanol CC(C)O. 
(\textbf{b}) Recombination: COCCO + CCOC(=O)C $\to$ CCOC.}
\label{fig:mutation-recombination}
\end{figure}

To make the mappings in Table~\ref{tab:chem-sda-operators} concrete, Figure~\ref{fig:mutation-recombination}a illustrates a \textit{mutation} event in which ethanol (CCO) is converted to isopropanol (CC(C)O), a local substitution that alters stability while preserving the scaffold.
Figure~\ref{fig:mutation-recombination}b shows a \textit{recombination} event: 
COCCO and CCOC(=O)C combine to yield CCOC, joining fragments from different
parents into a new motif. Here, recombination is performed in fragment space on symbolic molecular representations rather than as a stoichiometric chemical reaction; parent structures act as templates, and unused fragments are not interpreted as physical byproducts. These cases highlight the two main operator types:
mutations correspond to local functional modifications, while recombinations
generate novelty by linking distinct building blocks.

Note that some of these mutations and recombinations do not correspond to single-step reactions in real chemistry and, in practice, may require multistep synthesis pathways, catalysts, or additional energy input. In the SDA framework, however, they are abstracted into single operators in order to isolate the role of persistence bias and population dynamics, leaving mechanistic details to future, more chemically specific implementations.

\begin{figure}[H]
    \centering
    \includegraphics[width=0.7\textwidth]{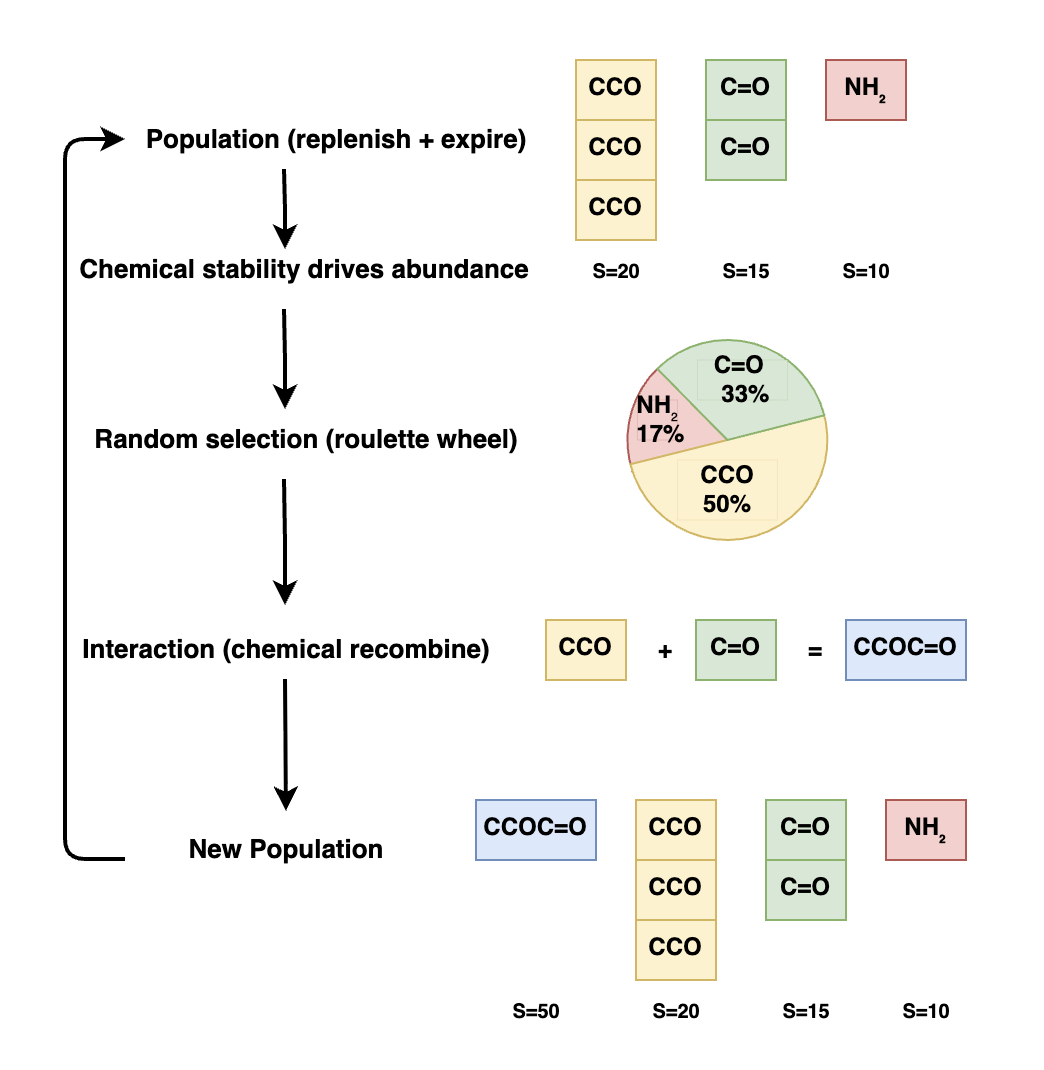}
    \caption{Schematic of the SDA/GA loop in chemical symbol space. 
    Each generation, motifs are replenished and expired, with stability values ($S$) 
    setting their relative abundance (e.g., ethanol-like \texttt{CCO}, $S=20$; carbonyl 
    \texttt{C=O}, $S=15$; amine \texttt{NH$_2$}, $S=10$). This skewed distribution biases 
    roulette-wheel selection toward more persistent motifs. Recombination events 
    (e.g., \texttt{CCO + C=O $\to$ CCOC(=O)}) produce more stable compounds 
    ($S=50$) that come to dominate the population. The loop illustrates how 
    persistence imbalances alone generate effective selection dynamics in chemical systems.}
    
    \label{fig:sda-chem-loop}
\end{figure}

This schematic illustrates the SDA/GA cycle instantiated in the chemical space. As in the symbolic ABC case, persistence bias amplifies the abundance of more stable motifs, which then dominate recombination events and seed increasingly complex products. The point is not to capture mechanistic detail but to show how stability values alone, when coupled with stochastic assembly and replenishment, are sufficient to generate skewed population structures and emergent selection dynamics. In this sense, chemical SDA/GA provides a bridge between abstract population models and realistic prebiotic chemistry.

\subsection{Stability and Environmental Drivers}\label{environ}

In SDA, stability $S(p)$ is defined as persistence between generations: the expected number of time steps that a pattern remains active in the population. This differs from conventional chemical usage, where stability refers to thermodynamic quantities or kinetic barriers. Instead, SDA abstracts these influences into a single phenomenological lifetime that governs continued participation in the generative process. $S(p)$ is then not an energy measure, but a measure of persistence informed by thermodynamics, kinetics, and environment.

The environmental drivers set these persistence times. Real chemical systems do not explore the assembly space uniformly: reaction outcomes depend on temperature gradients, redox, and pH cycles, porous flows, catalytic surfaces, and external energy inputs such as UV radiation or electrical discharges. These factors maintain systems far from equilibrium, making persistence differentials consequential. By contrast, laboratory chemistry is typically designed to reach equilibrium, after which dynamics cease. SDA focuses on the opposite regime: in open, driven systems, replenishment and fluctuation transform heterogeneous persistence into an evolutionary process.

Such conditions are documented in origin-of-life research. Hydrothermal vent systems provide sustained energy fluxes, catalytic mineral interfaces, and persistence gradients across thermal and redox boundaries \citep{martin2008energy}. Wet–dry cycling environments, including tidal flats and fluctuating hydrothermal pools. Similarly, couple-replenishment, concentration, and selective persistence through repeated dehydration–rehydration cycles \citep{damer2015coupled, rajamani2008lipid}. These environments do not specify particular outcomes, but generically realize the open, driven conditions under which persistence-driven selection operates. From this perspective, stability-as-persistence is not an alternative to chemical energetics but a population-level abstraction of their combined effects under environmental drive. Where stochastic assembly, replenishment, and differential persistence coexist, biased sampling and selection-like dynamics naturally emerge.

\section{Chemical SDA/GA Simulation}

\subsection{Methods}
To extend the symbolic SDA/GA framework into the chemical domain, we reused the same SDA simulation loop with two modifications. First, symbolic string elements were replaced with a small set of simple SMILES fragments representing hydrocarbons (e.g., methane, ethane, propane), oxygen- and nitrogen-containing groups (e.g., hydroxymethyl, amine, amino alcohol), and one aromatic ring (benzene), providing a rudimentary pool of prebiotic building blocks. The molecules were represented as SMILES strings and manipulated using the RDKit toolkit \cite{landrum2006rdkit}. Recombination was implemented through BRICS-based fragmentation and reassembly \cite{degen2008art}, ensuring valence plausibility and avoiding chemically impossible bonds. Mutation was modeled as single-site perturbations, analogous to copying errors or diffusion-induced reactions. These design choices follow established best practices for applying genetic algorithms to molecular discovery \cite{janet2023bestpractices}.

Stability values were estimated by an RDKit-based heuristic function $S(p)$ intended to approximate relative persistence under a neutral generative environment by mapping a continuous score $g(p)$, based on simple structural descriptors, to an integer lifetime:
\begin{equation}
S(p)=\mathrm{clip}_{[S_{\min},S_{\max}]}\!\left(\left\lfloor g(p)\right\rceil\right),
\end{equation}
where $\lfloor\cdot\rceil$ denotes rounding to the nearest integer. The score was defined as:
\begin{equation}
\begin{aligned}
g(p) &= 5 + 0.8\,\mathrm{HA}(p)
      + 1.0\,\mathrm{AR}(p)
      + 0.5\,\min\!\big(2,\mathrm{R}(p)\big)
      + 0.25\,\min\!\big(4,\mathrm{HBD}(p)+\mathrm{HBA}(p)\big) \\
&\quad - 0.3\,\min\!\big(10,\mathrm{Rot}(p)\big)
      - 1.0\,\min\!\big(2,\mathrm{FC}(p)\big)
      - 0.5\,\max\!\big(0,\mathrm{HA}(p)-30\big),
\end{aligned}
\end{equation}
where $\mathrm{HA}$: heavy atom count, $\mathrm{AR}$: aromatic rings, $\mathrm{R}$: aliphatic rings, $\mathrm{HBD/HBA}$: hydrogen-bond donors and acceptors, $\mathrm{Rot}$: rotatable bonds, and $\mathrm{FC}$: total formal charge magnitude. The resulting score was clipped to the range $[1,40]$ generations, ensuring bounded lifetimes while preserving large relative persistence differences.

In this way, fitness was not externally imposed but emerged from environmental stability constraints encoded through persistence, consistent with the SDA principle that selection arises from differential survival rather than from a predefined objective. All other aspects of the simulation loop remained unchanged: base fragments were replenished each generation, new compounds were formed by recombination and mutation, and expiration times were determined by $S(p)$. Thus, the chemical simulations can be understood as a direct extension of symbolic SDA, with minimal changes distinguishing abstract informational dynamics from chemically plausible ones.

\subsection{Results}

We ran chemical SDA simulations for 1000 generations with 200 interactions per generation and a replenishment rate of five. The results reveal how stability-driven persistence produces skewed population structures, motif-level evolution, and system-wide dynamics that parallel both genetic algorithms (GAs) and ecological systems.  

\begin{figure}[h]
\centering
\includegraphics[width=16cm]{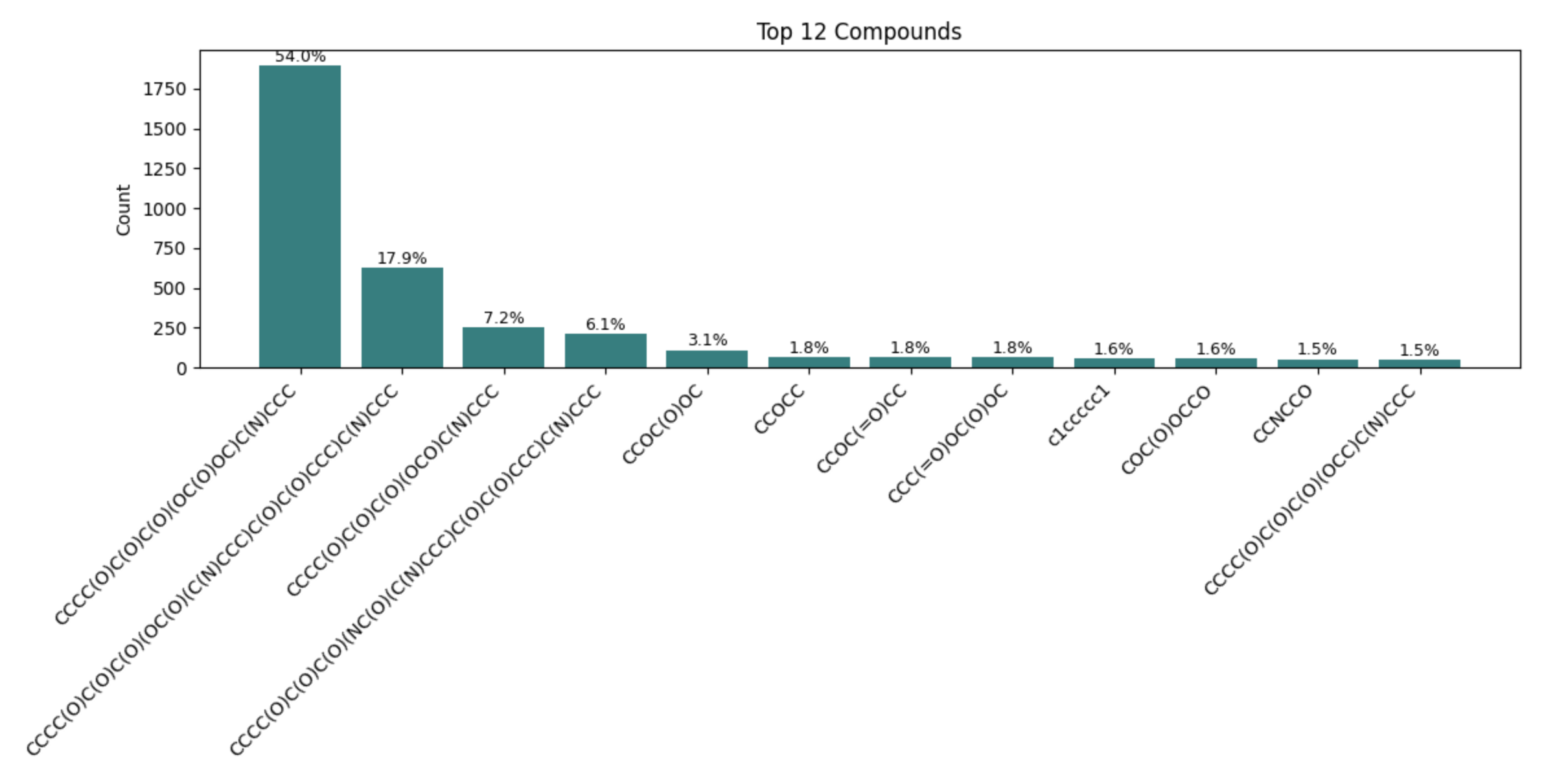}
\caption{Histogram of the twelve most frequent compounds at generation 1000, expressed as counts and percentages. One compound accounts for $\sim$54\% of the population, with the next runner-up at $\sim$18\%.}
\label{fig:chem-compound-hist}
\end{figure}

Figure~\ref{fig:chem-compound-hist} shows the distribution of the twelve most abundant compounds in generation 1000. A single ester-like motif (CCCC(O)C(O)C(O)(OC(O)OC)C(N)CCC) dominates more than half of the population ($\sim$54\%), while the second most frequent compound---a recursive oligomer in which the polyol core is extended by an additional C(N)CCC branch---accounts for 18\%. The third and fourth most frequent motifs (at 7\% and 6\%, respectively) are close structural relatives: one retains the O–C(=O)–O ester fragment in a simpler form, while the other incorporates an amide-like NC(O) substitution into the same polyol backbone. Together, these four scaffolds account for over 85\% of the entire pool.

From a GA perspective, this concentration illustrates how roulette-wheel selection emerges naturally from persistence: once a motif survives longer, it contributes disproportionately to the parent pool and thus amplifies its frequency. From a chemical perspective, the winners all share the same polyol–amine backbone, differing only in whether the substituent is ester-like, recursive, or amide-like. In other words, the system converges on a GA schema (the hydroxylated carbon scaffold with an appended amine) and explores variations within that schema, selecting for those with the greatest stability. The dominance of these few compounds therefore reflects both the GA principle of schema preservation and the chemical principle that stable substituents accumulate over evolutionary time.

\begin{figure}[h]
    \centering
    \includegraphics[width=1\textwidth]{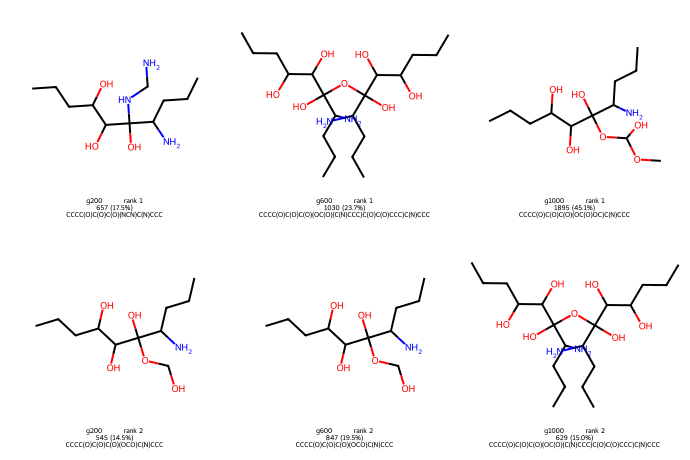}
    \caption{Evolution of the two top motifs across 200, 600, and 1000 generations. Early competition between aminomethylamino–substituted and ester-like variants gives way to long-term fixation of a single stable scaffold.}
    \label{fig:chem-top-evo}
\end{figure}

Figure~\ref{fig:chem-top-evo} traces the trajectories of the two dominant scaffolds.  
In generation 200, the population is divided between an aminomethylamino–substituted polyol (–NH–CH$_2$–NH$_2$; 17.5\%) and a related ester-like variant (14.5\%), both sharing a hydroxylated backbone with an appended amine chain.

By generation 600, a recursive oligomer elaborating the aminomethylamino–substituted branch expands to 23.7\%, while the ester-like motif remains at 19. 5\%. This phase reflects SDA’s tendency to preserve a common schema while generating more complex derivatives.  

By generation 1000, the simpler ester-like scaffold dominates nearly half of the pool (45. 1\%), while the recursive oligomer decreases to 15. 0\%. The trajectory illustrates schema competition: Multiple variants emerge from a shared template, but persistence imbalances ultimately favor the scaffold that best balances stability and generativity. Chemically, this shows how modest substituents (ester-like O–C(=O)–O groups) can outperform bulkier elaborations, driving convergence on motifs that are robust and reproductively generative.

\begin{figure}[h]
\centering
\subfloat[\centering Entropy dynamics]{
    \includegraphics[width=0.45\textwidth]{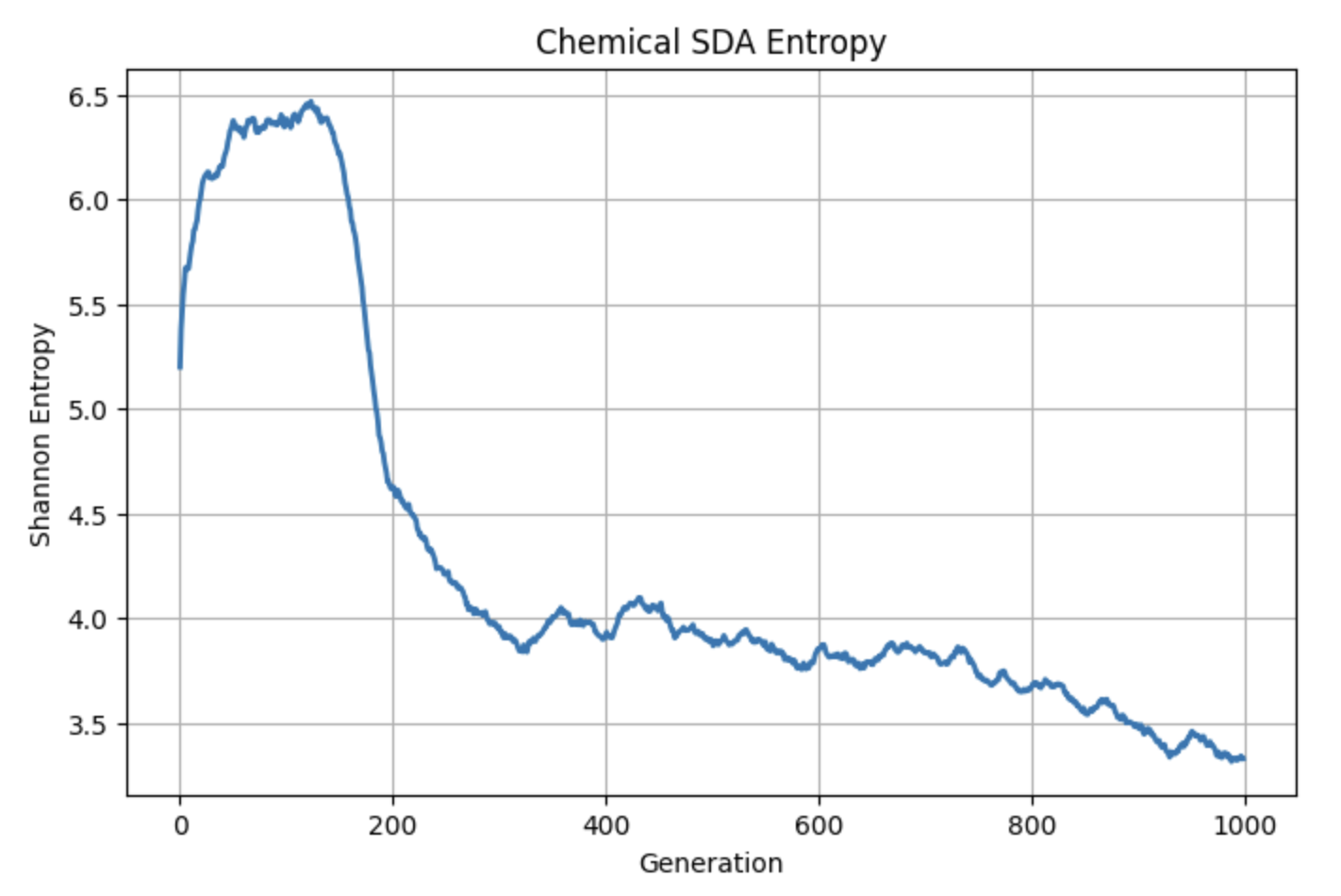}
    \label{fig:chem-entropy}
}
\hfill
\subfloat[\centering Diversity dynamics]{
    \includegraphics[width=0.45\textwidth]{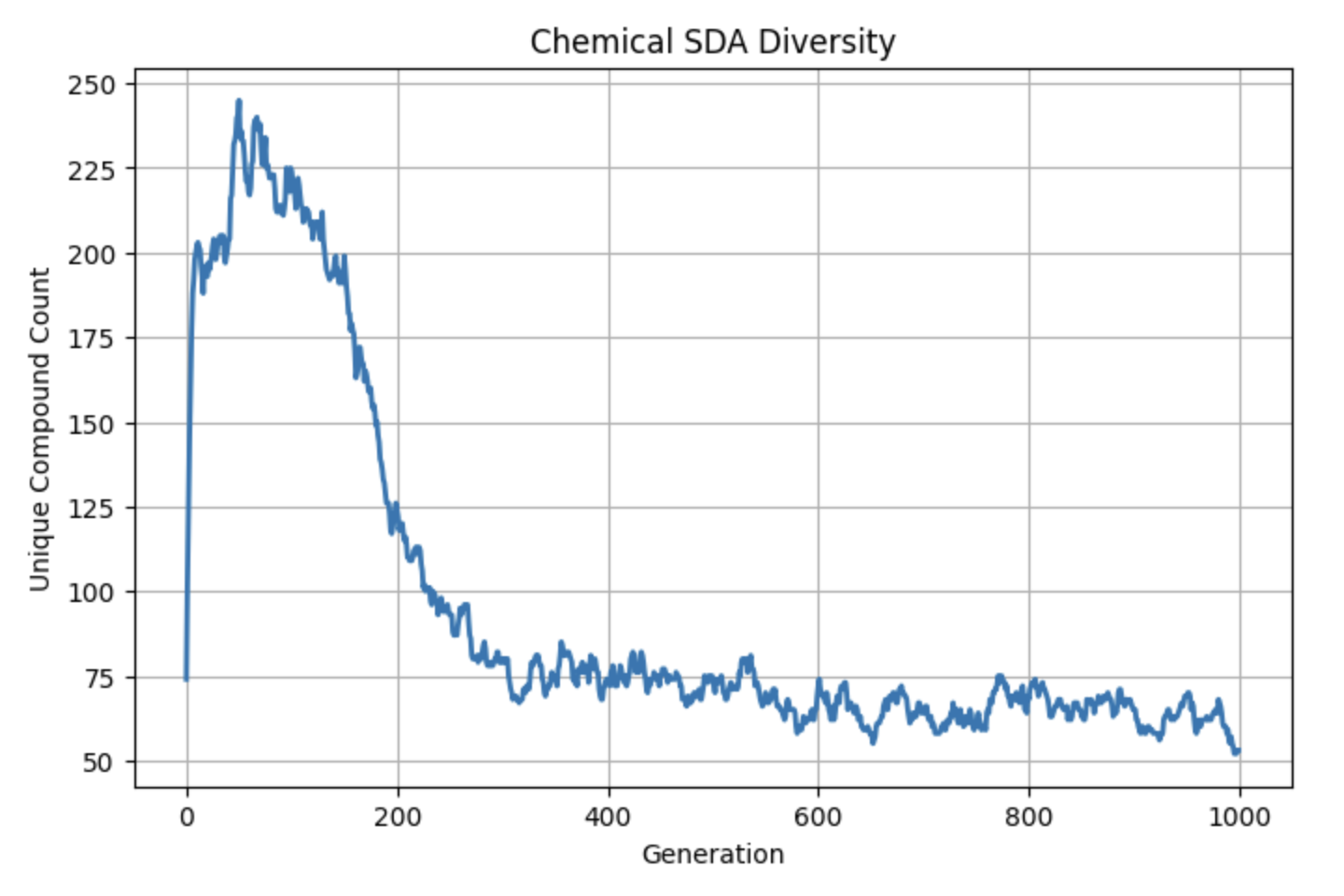}
    \label{fig:chem-diversity}
}
\caption{System-level traces of chemical SDA. (\textbf{a}) Shannon entropy rises sharply as new compounds appear, then declines steadily as dominant motifs consolidate. (\textbf{b}) Unique compound diversity follows a parallel trajectory, peaking at over 200 species before collapsing to $\sim$50 by the end of the run.}
\label{fig:chem-entropy-diversity}
\end{figure}

The entropy and diversity dynamics, shown in Figure~\ref{fig:chem-entropy-diversity}, both trace the consolidation of the system. During the first 100–200 generations, entropy increases and diversity expands to more than 200 distinct compounds as exploration dominates. Thereafter, both measures decline in parallel: entropy falls steadily, while diversity collapses to about 50 species by generation 1000, with the vast majority of the population concentrated in just two scaffolds. This concentration does not indicate convergence to an optimal solution or a terminal state; rather, it reflects a transient dominance shaped by current stability conditions. Together, these trends show how stability-driven persistence prunes the search space, channeling the system toward a narrowed set of long-lived motifs while remaining open to the emergence of new motifs that can, at any time, displace existing scaffolds and reshape the population.

\begin{figure}[h]
\centering
\subfloat[\centering Novelty over time]{
    \includegraphics[width=0.45\textwidth]{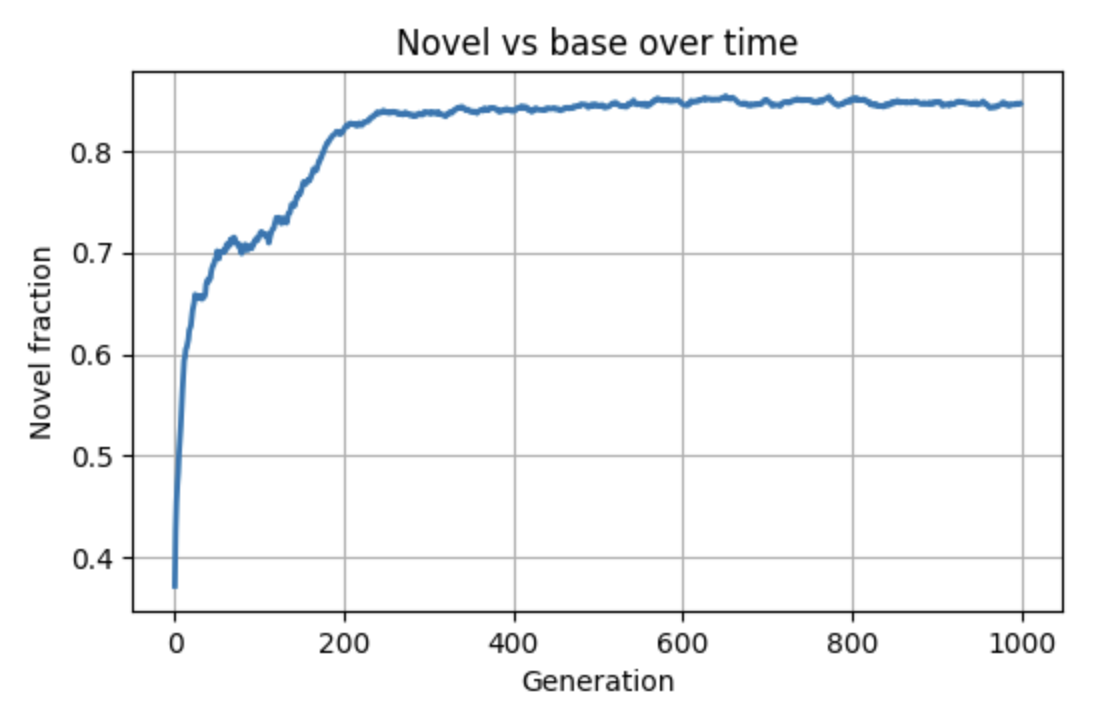}
    \label{fig:chem-novel}
}
\hfill
\subfloat[\centering Rank--abundance distribution]{
    \includegraphics[width=0.45\textwidth]{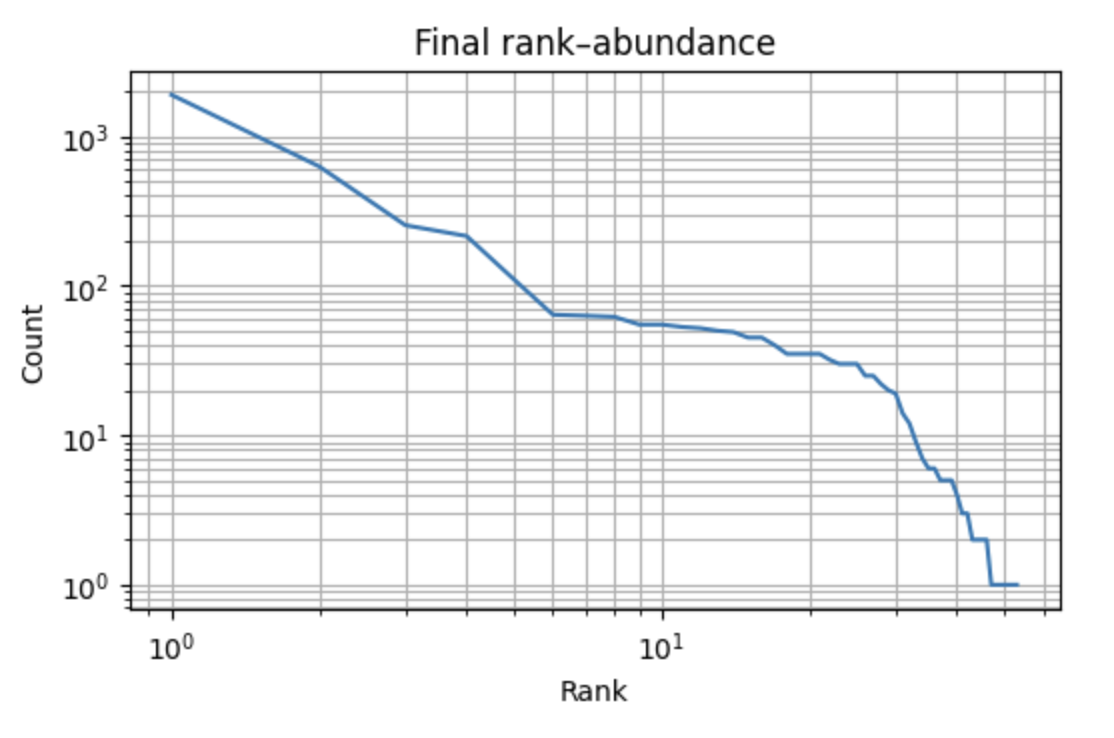}
    \label{fig:chem-rank}
}
\caption{Novelty and abundance structure in chemical SDA. (\textbf{a}) Fraction of novel compounds (not present in the initial fragment pool) over time. Novelty rapidly overtakes the base pool and stabilizes above 80\%. (\textbf{b}) Rank--abundance distribution at generation 1000 (log--log scale), showing a heavy-tailed form where a few motifs dominate while many persist at low frequency.}
\label{fig:chem-novel-rank}
\end{figure}

Figure~\ref{fig:chem-novel-rank} summarizes the resulting novelty and abundance structure. The fraction of novel compounds rises rapidly, exceeding 80\% by generation 200 and remaining high thereafter, indicating sustained exploration driven by ongoing recombination rather than recycling of the initial pool. At the same time, the population exhibits a heavy-tailed rank–abundance distribution, with a small number of dominant motifs and a long tail of rare variants. Together, these patterns show that SDA dynamics couple open-ended novelty with structured, law-like population organization.

\subsection{Interpretation}

These results show that chemical SDA/GA realizes the dynamics of a natural genetic algorithm, with population skew arising spontaneously from persistence imbalances: compounds that survive longer contribute disproportionately to the parent pool, yielding fitness-proportional sampling. This behavior is visible in the compound histograms and motif trajectories (Figures~\ref{fig:chem-compound-hist}, \ref{fig:chem-top-evo}), which reveal scaffold-level dominance and competition.

System-level traces (Figures~\ref{fig:chem-entropy}--\ref{fig:chem-novel}) explain how this structure forms. Entropy and diversity initially expand but later contract as persistence narrows the space of viable motifs, while novelty remains high as recombination continually introduces new variants. The resulting rank–abundance distribution (Figure~\ref{fig:chem-rank}) follows heavy-tailed statistical patterns characteristic of complex adaptive systems.

Even with a heuristic stability function, the most abundant compounds exhibit chemically plausible motifs. Their persistence reflects robustness under continual turnover rather than optimization toward a predefined target, supporting the SDA/GA hypothesis that stability-driven persistence with recombination and mutation generates sustained novelty and structured population-level order.

\subsection{Equilibrium Models versus Selection-Driven Evolution}

It is useful to contrast these results with equilibrium-based analyses such as those derived from mass–action kinetics (MAK) models \cite{fogler1999chemical,TuranyiTomlin2014}. To render the governing equations linear and analytically tractable, such models typically assume constant reaction rates and well-mixed dynamics. These assumptions ensure mathematical stability but suppress the very mechanisms that enable open-ended evolution. With constant rates, all species participate equivalently in a Markovian flow, and long-term behavior is governed by equilibrium concentrations determined by stoichiometry rather than by differential persistence. Once equilibrium is reached, directional change ceases.

In contrast, the SDA/GA framework does not assume constant rates or equilibrium conditions. As established in the population update of Section~\ref{sec:pattern-evolution}, effective turnover depends on stability and feeds back on the evolving population distribution: persistence times bias which patterns participate in further interactions, which in turn shape the population. This nonlinear, self-consistent feedback creates selection: stable motifs accumulate, unstable ones vanish, and parent sampling becomes increasingly biased. The simulations reflect this mechanism directly: entropy and diversity initially expand, then contract as a few stable motifs dominate, while recombination continually introduces novelty and yields heavy-tailed population structure. What appears in equilibrium theory as a static steady state is, in the SDA view, a transient outcome of ongoing competition, turnover, and selection. This distinction clarifies why equilibrium approaches are limited in their ability to capture chemical systems that exhibit genuinely evolutionary dynamics.

\subsection{Contrast to Supervised Learning and Genetic Programming}

Chemical SDA/GA is conceptually distinct from supervised learning and conventional GA/GP used in drug discovery. In supervised machine learning, search is explicitly directed toward minimizing a predefined loss function, while in many GA and GP implementations selection is guided by a user-specified fitness function encoding a target solution. In contrast, chemical SDA/GA has no external objective or target pattern. Selection arises solely from persistence: molecules that survive longer contribute more offspring, and the evolving chemical environment continuously reshapes which motifs are viable.

In this sense, SDA/GA resembles a form of reinforcement learning without an externally defined reward signal. Persistence functions as an intrinsic feedback mechanism rather than an engineered objective, and adaptation emerges from the interaction between stochastic assembly and environmental constraints. The resulting dynamics are open-ended: populations drift toward increasingly stable and better-fitting motifs not because a solution was specified in advance, but because stability differentials bias exploration in a continually changing chemical landscape.

\subsection{Emergence of a Natural Genetic Algorithm}

\begin{figure}[h]
    \centering
    \includegraphics[width=0.6\textwidth]{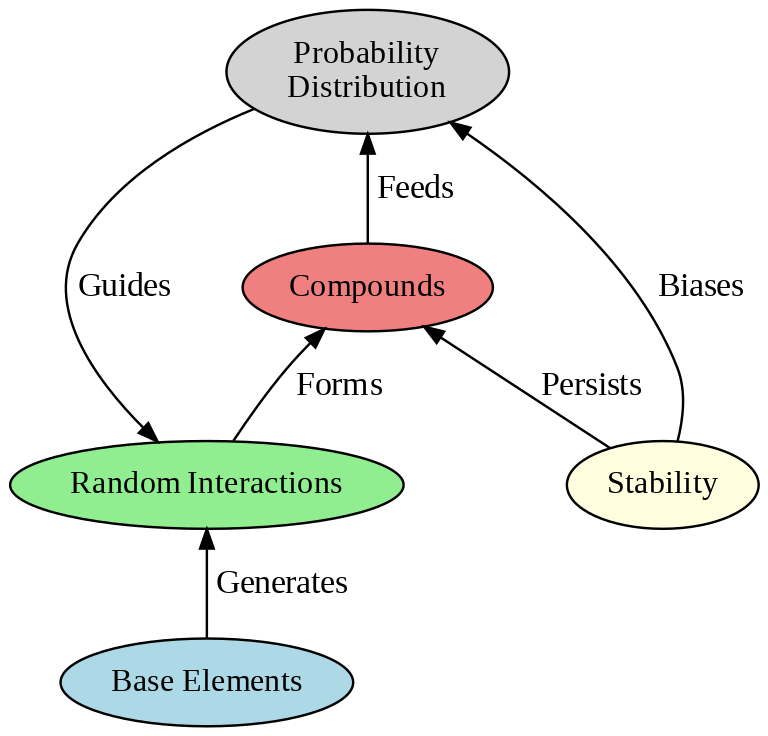}
    \caption{Feedback structure underlying chemical SDA dynamics. Base elements interact randomly to form compounds. Compounds that persist contribute more to the population, biasing the probability distribution. This distribution in turn guides future interactions, closing a feedback loop. The result is emergent roulette-wheel selection.}
    \label{fig:top-down}
\end{figure}

The feedback structure shown in Figure~\ref{fig:top-down} provides a mechanistic explanation for how the GA-like search spontaneously emerges in chemical SDA. In conventional genetic algorithms, roulette-wheel selection is imposed externally: the algorithm samples parents in proportion to a programmer-defined fitness function. In SDA, no such programmer is required. Instead, stability imbalances ensure that some compounds persist longer than others. Persistence in turn biases the probability distribution over the population, which then guides future interactions. 

This feedback loop closes the causal chain: the compounds shape the distribution, the distribution shapes sampling, and sampling determines which compounds are likely to interact. In effect, the selection pressure is not imposed top-down, but emerges bottom-up from the persistence differences within the population. The result is indistinguishable from roulette-wheel selection, yet it arises naturally from the dynamics of the system.

This perspective also clarifies the often misused notion of top-down causation \cite{noble_dance}. There is no need to posit a mystical programmer or abstract force directing the search. What appears as a top-down influence of the probability distribution on individual elements is mechanistically explained by the accumulation of persistence imbalances at the population level. Stability alone suffices to generate the feedback necessary for open-ended, GA-like evolution.

\section{The Evolutionary Ladder Hypothesis}  

\begin{figure}[h]
\centering
\includegraphics[width=14cm]{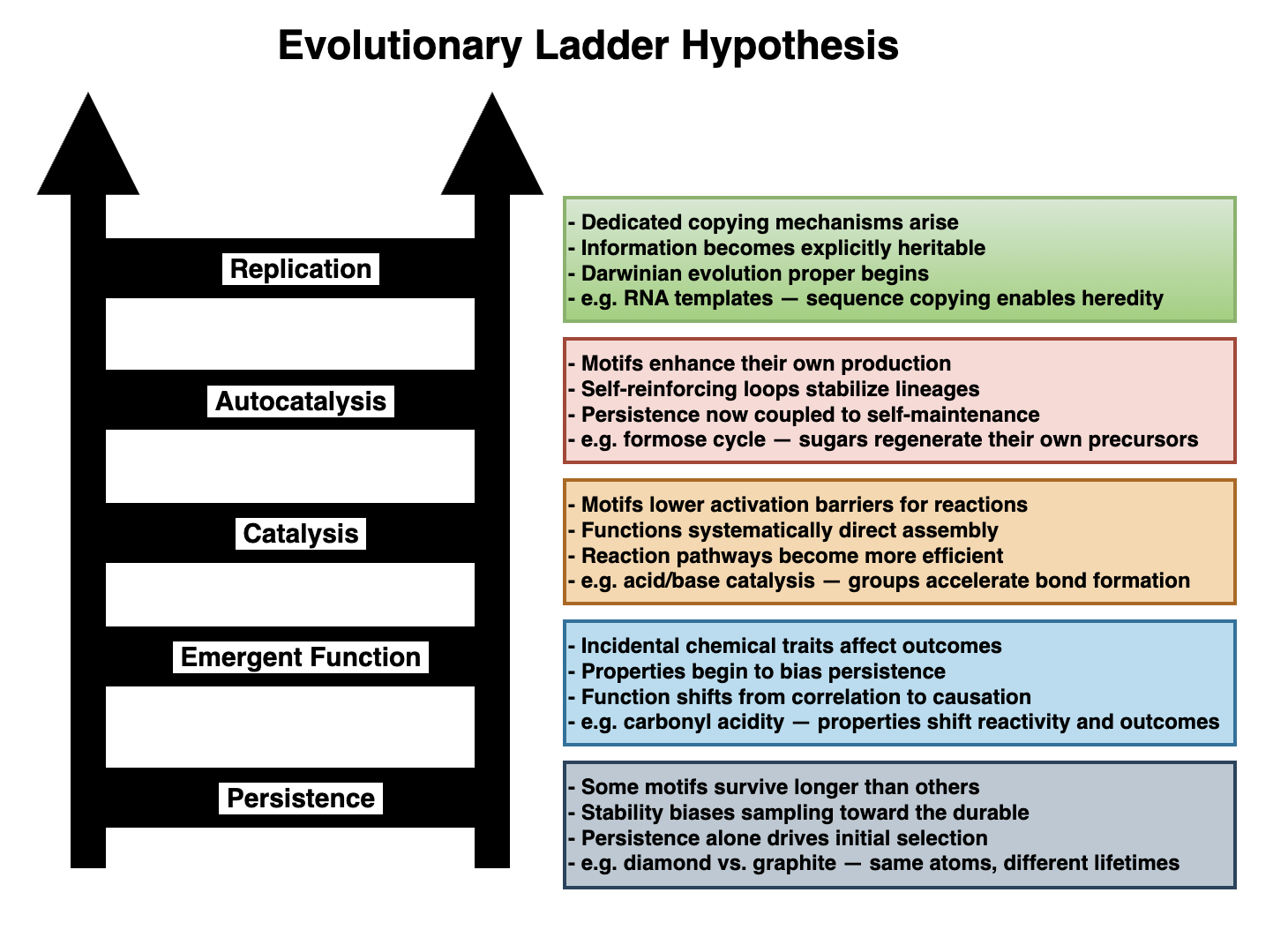}
\caption{\textbf{SDA Evolutionary Ladder.} A stepwise progression from persistence-only dynamics to Darwinian replication. \emph{Persistence}: some motifs are longer-lived, biasing sampling; 
\emph{Emergent Function}: incidental chemical properties begin to bias persistence and can become causal; 
\emph{Catalysis}: functional motifs systematically lower barriers and direct assembly; 
\emph{Autocatalysis}: motifs enhance their own production, forming self-reinforcing loops; 
\emph{Replication}: sequence-specific copying makes information explicitly heritable. 
The boxes at right give concise examples for each rung (e.g., diamond vs.\ graphite; carbonyl acidity; acid/base catalysis; the formose cycle; RNA templating). 
Colors encode increasing organizational complexity.}
\label{fig:sda-ladder}
\end{figure}

Figure~\ref{fig:sda-ladder} illustrates the evolutionary ladder hypothesis, a proposed continuum linking chemistry to biology. At the base lies \emph{persistence}: Some motifs simply last longer than others. Diamonds outlast graphite under pressure, and stable scaffolds remain in circulation while unstable ones vanish. In this phase, persistence alone biases sampling, creating the first form of selection without function.  

The next rung, \emph{emergent function}, arises when incidental chemical traits begin to influence persistence. For example, a carbonyl motif that mutates into a carboxylic acid changes acidity, altering how it interacts with the environment. In SDA terms, this is a single mutation, but chemically it shifts reactivity, allowing the motif to persist in contexts where neutrality would not. At this stage, properties are still largely correlated with persistence, but they begin to bias outcomes systematically.  

A concrete example illustrates the transition from correlation to causation. Consider a carbonyl mutating into a carboxylic acid. The acid group can protonate bases and lower activation barriers, thereby catalyzing an unrelated condensation reaction that produces a more persistent ester or amide motif. In SDA terms, the first motif’s persistence is no longer merely correlated with stability: its functional effect causally increases the survival and reproduction of another motif. The second motif then increases in frequency not only due to its intrinsic stability, but because the acid \emph{phenotypic function} enabled its production. Function thus feeds directly into the feedback loop of persistence, transforming passive traits into causal drivers of selection.  

This transition naturally leads to \emph{catalysis}, where the motifs systematically lower the activation barriers, channeling reaction pathways toward more efficient outcomes. Once catalysis is established, some motifs go further, entering \emph{autocatalysis}. In this rung, motifs not only stabilize others, but reinforce their own production. The formose sugar cycle, for example, regenerates its own intermediates, creating a self-reinforcing lineage. Persistence is now coupled to self-maintenance.  

Finally, \emph{replication} \cite{england2013statphys} emerges when the copying mechanisms become explicit and sequence-specific. At this rung, information is no longer implicit in persistence alone, but is encoded in heritable structures such as RNA templates. Darwinian evolution proper begins only here, when variation, inheritance, and selection operate together.  

Seen as a continuum, replication is therefore not the foundation of evolution but one rung in a broader ladder that begins with persistence. Our simulations capture the earliest stages, showing that persistence imbalances alone can launch evolutionary trajectories, with phenotypic functions serving as the bridge from chemistry to biology.

\subsection{Generic Conditions for Persistence-Driven Selection}\label{sec:generic}

The ladder hypothesis above presupposes that the bottom rung (differential persistence) exists wherever chemistry runs under open conditions. This is not an additional assumption, but a generic consequence of the physical structure. Persistence imbalances in real chemistry span many orders of magnitude: binding energies, activation barriers, structural reinforcement, and hierarchical assembly produce lifetimes ranging from transient excited states to structures stable on geological timescales. These differentials are not incidental but follow from basic features of matter: symmetry, energetic minima, geometric and steric constraints, and the recursive assembly of stable subunits into more stable composites.

Uniform persistence across all motifs would require the cancellation of these effects across the entire combinatorial space, an exceptionally tuned configuration that no realistic chemistry exhibits. Uneven persistence landscapes are generic: any physical realization of an open chemical system will exhibit some motifs that outlast others, and once such differentials exist, the SDA feedback of stability $\to$ persistence $\to$ population skew $\to$ biased sampling proceeds automatically. Persistence-driven selection therefore requires neither special initial conditions nor parameter tuning; it requires only that chemistry be embedded in an open, replenished system far from equilibrium: hydrothermal vents, wet--dry cycling environments, and other prebiotic settings discussed in Section~\ref{environ}.

Viewed this way, the bottom rung of the evolutionary ladder is not a narrow or contingent configuration but a default feature of nonequilibrium chemistry. The substantive question is therefore not whether stability-driven assembly can occur, but how far up the ladder it ascends in any given environment.

\section{Conclusions}

This work develops \textit{Stability-Driven Assembly} (SDA) as a mechanism by which selection emerges without explicit fitness functions or replicators. Persistence imbalances skew population composition, and this skew feeds back into sampling so that longer-lived motifs are preferentially reused. The resulting loop of \emph{create $\rightarrow$ persist $\rightarrow$ sample} implements roulette–wheel selection, achieving a natural genetic algorithm (SDA/GA) in which fitness is supplied by the environment, in the form of persistence, rather than externally imposed.

Chemical simulations demonstrate these dynamics in practice. Populations become dominated by a small number of stable scaffolds, entropy and diversity initially expand and then contract, and rank–abundance distributions develop heavy tails with few dominant motifs and many rare ones. These signatures reflect selection acting on a generative process: novelty is continually produced, but only persistent structures accumulate.

This behavior contrasts sharply with equilibrium approaches, such as constant-rate mass–action kinetik (MAK) models, which suppress this feedback by construction. In SDA/GA, effective rates are heterogeneous because persistence differs between motifs, so the population never relaxes to a fixed equilibrium. Instead, the distribution itself evolves under ongoing competition and turnover.

The present formulation remains intentionally simplified, relying on heuristic stability functions, abstracted operators, well-mixed dynamics, and the omission of solvent and kinetic detail. Future work should extend these models with multi-run statistics, richer stability functions grounded in thermodynamics and kinetics, additional reaction classes, and experimental tests in open, driven reactors where persistence can be directly measured.

Viewed through SDA/GA, information is not a static property of structures but an emergent consequence of population dynamics. Regularities accumulate because persistence biases which motifs endure and recombine, so information itself evolves through biased exploration. From this perspective, order, information, and evolution are unified: stable motifs persist longer, recur more often, and proliferate through persistence-driven selection, pointing to a common physical principle linking chemistry, computation, and biology.





\printbibliography

\section*{Appendix}
\appendix

\section{Simulation Code}

The code and simulations corresponding to the figures and experiments presented in this
paper are available in this Google Colab notebook: \url{https://colab.research.google.com/drive/12wW42Fl2Zkopgicy59_JHHpap3vNXep6}
\end{document}